%% file: main.tex
\documentclass[conference]{IEEEtran}
\IEEEoverridecommandlockouts
\usepackage{cite}
\usepackage[caption=false,font=normalsize,labelfont=sf,textfont=sf]{subfig}
\usepackage{mathtools}
\usepackage[noend]{algpseudocode}
\usepackage[ruled ,vlined]{algorithm2e}

\usepackage{graphicx}
\usepackage{xspace}
\usepackage{color,soul}
\usepackage{amssymb}
\usepackage{booktabs}
\usepackage{makecell}
\usepackage{hyperref}
\usepackage{multirow}
\usepackage{pifont}
\usepackage{wrapfig}
\usepackage{svg}
\usepackage{float}
\usepackage{graphicx}
\usepackage{float}
\usepackage{stfloats}
\usepackage{environ}
\usepackage{tikz}
\usepackage{svg}
\usepackage{amsthm}

\NewEnviron{elaboration}{
\par
\begin{tikzpicture}
\node[rectangle,minimum width=0.3\textwidth] (m) {\begin{minipage}{0.3\textwidth}\BODY\end{minipage}};
\draw[dashed] (m.south west) rectangle (m.north east);
\end{tikzpicture}
}

\def\BibTeX{{\rm B\kern-.05em{\sc i\kern-.025em b}\kern-.08em
    T\kern-.1667em\lower.7ex\hbox{E}\kern-.125emX}}

\newtheorem{definition}{Definition}

\newtheorem{example}{Example}[section]

\newcommand{\approach}{{Hybrid-CAR}\xspace}
\newcommand{\tool}[1]{{\sc #1}}
\newcommand{\CAR}{CAR\xspace}
\newcommand{\BMC}{BMC\xspace}
\newcommand{\IMC}{IMC\xspace}
\newcommand{\PDR}{PDR\xspace}
\newcommand{\SAT}{SAT\xspace}

\newcommand{\UC}{UC\xspace}
\newcommand{\UCs}{UCs\xspace }

\newcommand{\simplecar}{\tool{SimpleCAR}\xspace}

\newcommand{\local}{CoreLocality\xspace} 
\newcommand{\tline}{\Xhline{3\arrayrulewidth}}

\newcommand\os[1]{\textcolor{blue}{~[#1]}} 
\newcommand\yb[1]{\textcolor{purple}{~[#1]}} 
    
\begin{document}

\title{Revisiting Assumptions Ordering in CAR-Based Model Checking}



\author{\IEEEauthorblockN{Yibo Dong\IEEEauthorrefmark{2},
Yu Chen\IEEEauthorrefmark{4}, 
Jianwen Li
\IEEEauthorrefmark{3}, 
Geguang Pu\IEEEauthorrefmark{3}, 
Ofer Strichman\IEEEauthorrefmark{5}
}
\IEEEauthorblockA{\IEEEauthorrefmark{2}Each China Normal University, China, \href{mailto:}{prodongf@gmail.com}\\
\IEEEauthorrefmark{3}Each China Normal University, China, \href{mailto:xxx@xxx}{\{jwli,ggpu\}@sei.ecnu.edu.cn}\\
\IEEEauthorrefmark{4}Chuzhou University, China,  \href{mailto:}{chenyu@chzu.edu.cn}}
\IEEEauthorrefmark{5}Technion, Isreal,  \href{mailto:}{ofers@technion.ac.il}
}

\maketitle

\begin{abstract}
Model checking is an automatic formal verification technique that is widely used in hardware verification. The state-of-the-art complete model-checking techniques, based on IC3/PDR and its general variant CAR, are based on computing symbolically sets of under - and over-approximating state sets (called `frames') with multiple calls to a SAT solver. The performance of those techniques is sensitive to the order of the assumptions with which the SAT solver is invoked, because it affects the \emph{unsatisfiable cores} --- which the solver emits when the formula is unsatisfiable --- that crucially affect the search process. This observation was previously published in~\cite{DLPVR19}, where two partial assumption ordering strategies, \emph{intersection} and \emph{rotation} were suggested (partial in the sense that they determine the order of only a subset of the literals). In this paper we extend and improve these strategies based on an analysis of the reason for their effectiveness. We prove that intersection is effective because of what we call \emph{locality} of the cores, and our improved strategy is based on this observation. We conclude our paper with an extensive empirical evaluation of the various ordering techniques. 
One of our strategies, \approach, which switches between strategies at runtime, not only outperforms other, fixed ordering strategies, but also outperforms other state-of-the-art bug-finding algorithms such as \tool{ABC-BMC}. 

\end{abstract}


\input{1.introduction}

\input{2.preliminary}
\input{3.revisit_prior_work}

\input{4.new_approaches}

\input{5.experiment }

\input{6.conclusion }

\clearpage
\bibliographystyle{plain}
\bibliography{main}

\input{bio}
\end{document}

%% file: 1.introduction.tex
\section{Introduction}

\emph{Model checking} is an automatic formal verification technique that is central in the hardware design community~\cite{BES16,JM09}. Given a model $M$ and a temporal property $P$ over its variables, it checks whether all the behaviours of $M$ satisfy $P$, i.e., whether $M\models P$. 
Once a system behaviour is detected to violate $P$, the model checker returns a \emph{counterexample} as the evidence, which demonstrates the execution of the system leading to the property violation. Such a process is called \emph{bug-finding}. If $P$ is a \emph{safety} property, the violation of $P$ is witnessed by a counterexample made of a finite number of states. It is well known that model checking on safety properties can be reduced to reachability analysis~\cite{BCMDH90}. 

State-of-the-art safety model checking techniques include Bounded Model Checking (\BMC) \cite{BCCFZ99,BCCZ99}, Interpolation Model Checking (\IMC) \cite{McM03}, Property Directed Reachability (\PDR) (also called IC3)~\cite{Bra11,EMB11}, and Complementary Approximate Reachability (CAR) \cite{LZZPV17}, all of which integrate a SAT solver internally. \BMC is an incomplete method (it is only used for finding bugs, not proving their absence) and is empirically very fast at finding relatively shallow bugs (i.e., after a relatively small number of steps from the initial state). \IMC, \PDR and \CAR are complete but are generally not as fast as \BMC in shallow bug-finding, and none of the existing implementations of those techniques dominates the other. In \cite{LZZPV17,LDPRV18}, it was empirically shown that within a given time and hardware resources, \CAR is able to solve unsafe instances that \BMC cannot, and safety instances that \IMC and \PDR cannot, while the converse is true as well. Therefore, a portfolio consisting of different techniques is often maintained for different verification tasks. 
However, model-checking (a PSPACE problem), always falls short of the performance needs in the industry when it comes to verifying large designs~\cite{TM11,DZYCCMLP23}. Indeed, performance optimisation of SAT-based model-checkers is an active research area. Some recent examples are \cite{ZXLPS22},\cite{xia2023igood}, ~\cite{RKSNCV23} and~\cite{SSCRW22}.

In this paper, we focus on improving the performance of \CAR. We will describe in detail how \CAR works in Section~\ref{sec:CAR}. It has many similarities to \PDR, which is better known, but also several distinctive features. For now, let us just mention that, similar to \PDR, it relies on many SAT calls over relatively easy formulas. One of its elements is a sequence of formulas $O_1\ldots O_k$, called the over-approximating frames (or \emph{$O$-frames}, for short), where $O_i$, $1\leq i \leq k$, over-approximates the states that can reach $\neg P$ within $i$ steps. \CAR gradually makes these frames more precise, i.e., less over-approximating, by removing from them states that cannot reach $\neg P$ within the given number of steps. One of the critical elements of this process is \emph{generalisation}, that is, the ability to remove many such states at once. This is done by finding the \emph{unsatisfiable core} (\UC) of unsatisfiable SAT calls. The research we report here focuses on improving the quality of those \UCs, which accelerates the narrowing process of the $O$-frames. To explain our contributions, let us briefly recall how modern SAT solvers find \UCs. 

The input to every \SAT call in \CAR (and \PDR) takes the form of $\bigwedge_{l\in\mathcal{A}}l\wedge \phi$, where $\phi$ is a Boolean formula in Conjunctive Normal Form (CNF) and $\mathcal{A}$ consists of a sequence of literals, called the \emph{assumptions}. Almost all modern CDCL-based \SAT solvers as of  \tool{Minisat}~\cite{ES03} and \tool{Glucose}~\cite{glucoseSATSolver, glucoseIncremental} support assumptions. They position the literals in $\mathcal{A}$, in order, as their first decisions, and perform Boolean Constraint Propagation (BCP) as usual. Unsatisfiability is detected when the BCP of an assumption contradicts the value of another literal (because recall, all the literals in $\mathcal{A}$ and those that are implied by them via BCP are implied by the formula regardless of any decision). By analysing the trail, the solver can detect which of the assumptions contributed to the conflict and emit this list of assumptions as the \UC, which is essentially a compact \emph{reason} for the unsatisfiability. In other words, the \UC is a subset of $\mathcal{A}$ that is sufficient for making $\phi$ unsatisfiable. 

There can be multiple {\UC}s in a given unsatisfiable formula, and the order of the assumptions may affect the \UC that is found (and this, in turn, affects the overall performance of the model-checker, whether it is \CAR or \PDR). More explicitly, the literals that are propagated earlier are more likely to appear in the returned \UC. Indeed, prior work leveraged this phenomenon to improve performance. Specifically, the IC3ref model checker~\cite{ictref}, which implements the original IC3 algorithm, sorts the literals in $\mathcal{A}$ in descending order based on their appearance frequencies. The \simplecar model-checker~\cite{li2018simplecar}, which implements \CAR, uses two different literal-ordering strategies, as reported in~\cite{DLPVR19}:  \emph{Intersection}, which prioritizes literals that are both in the current state and the latest generated \UC, and \emph{Rotation}, which prioritizes literals that are present in all previously explored states. This makes these literals more likely to appear as part of the generated \UC, and empirically improves bug-finding performance. Indeed, \simplecar is one of the baseline implementations against which we compare our contributions. In addition, we compare ourselves against the best known \BMC implementation, as well as previous optimizations that were applied to \CAR~\cite{ZXLPS22}. 

Our contributions are: 
\begin{enumerate} 
    \item We revisit one of the two heuristics proposed in \cite{DLPVR19}, called \emph{Intersection}, and suggest an explanation for its effectiveness. Briefly, we show that it leads to finding proofs of unsatisfiability faster because of what we call the \emph{locality} of the cores. Based on this observation, we propose an extension of this technique that improves locality, and decides on the order of more literals comparing to the original version of \emph{Intersection}. We also show how it affects a combination of \emph{locality} with another strategy from~\cite{DLPVR19} called \emph{Rotation}. 
    Our experimental results indicate that this leads to faster convergence and increases the number of cases that can be solved within a given timeout.
   \item We define the \emph{conflict literal} as the last literal found to be in the \UC by the SAT solver, and observe that unlike the other literals in the core, it is \emph{necessary} (without it, the other literals do not form a core). We show that by prioritizing these literals, proofs are found faster. 
   \item We suggest a method called \approach, which swaps \emph{\local}'s configurations during the search (based on giving a time-limit to each configuration), which not only outperforms the previous `static' ordering approaches but also any other bug-finding techniques, including \tool{ABC-BMC}, the state-of-the-art \BMC implementation~\cite{BM10} and previous published versions of \CAR.
  
    \item We provide an extensive empirical study about the influence of assumption ordering on the \UC generated, with the existing and new strategies, thereby empirically demonstrating the significance of literal ordering in \CAR-based model checkers.     
\end{enumerate}

We continue with preliminaries in the next section. 
In Section~\ref{sec:revisit} we describe the prior work of~\cite{DLPVR19} and explain why literal ordering matters. Section~\ref{sec:newapproach} describes our contribution, and Section~\ref{sec:experiments} describes the results of our empirical evaluation. Our conclusions are summarised in Section~\ref{sec:conclusion}.

%% file: 2.preliminary.tex
\section{Preliminaries}\label{sec:preliminary}

\subsection{Boolean Transition System}\label{preliminary:A}
A Boolean transition system $Sys$ is a tuple $(V, I, T)$, where $V$ and $V'$ denote the set of variables in the present state and the next state, respectively. The state space of $Sys$ is the set of possible variable assignments. $I$ is a Boolean formula corresponding to the set of initial states, and $T$ is a Boolean formula over $V\cup V'$, representing the transition relation. 
State $s_{2}$ is a successor of state $s_{1}$ iff $s_{1} \cup s_{2}' \models \emph{T}$, which is also denoted by $(s_1,s_2)\in T$.
A \emph{path} of length $k$ is a finite state sequence $ s_{1}, s_{2}, \dots, s_{k} $, where $(s_{i},s_{i+1})\in T $ holds for $(1\le i \le k -1)$. A state $t$ is reachable from $s$ in $k$ steps if there is a path of length $k$ from $s$ to $t$.
Let $X \subseteq 2^{V}$  be a set of states in \emph{Sys}. We denote the set of successors of states in $X$ as $R(X) = \{t \mid (s,t) \in T, s \in X\}$. Conversely, we define the set of predecessors of states in $X$ as $R^{-1}(X) = \{s \mid (s,t) \in T, t \in X\}$. Recursively, we define $R^{0}(X) = X$ and $R^{i}(X) = R(R^{i-1}(X))$ where $i \ge 0$, and the notation $R^{-i}(X)$ is defined analogously. In short, $R^{i}(X)$ denotes the states that are reachable from $X$ in $i$ steps, and $R^{-i}(X)$ denotes the states that can reach $X$ in $i$ steps.

\subsection{Safety Model Checking and Reachability Analysis}

Given a transition system  $Sys=(V, I, T)$ and a safety property $P$, which is a Boolean formula over $V$, a model checker either proves that $P$ holds for any state reachable from an initial state in $I$, or disproves $P$ by producing a \emph{counterexample}. In the former case, we say that the system is safe, while in the latter case, it is unsafe.
A counterexample is a finite path from an initial state $s$ to a state $t$ violating $P$, i.e., $t \in \neg P$, and such a state is called a \emph{bad} state.
In symbolic model checking, safety checking is reduced to symbolic reachability analysis. Reachability analysis can be performed in forward or backward search. Forward search starts from initial states $I$ and searches for reachable states of $I$ by computing $R^{i}(X)$ with increasing values of $i$, while backward search begins with states in $\neg P$ and computes 
$R^{-i}(X)$ with increasing values of $i$ to search for states reaching $\neg P$. Table~\ref{tab:Forward/Backward Search} gives the corresponding formal definitions.

\begin{table}[ht]
\renewcommand{\arraystretch}{1.4}
\caption{Standard reachability analysis.}
\label{tab:Forward/Backward Search}
\centering
\begin{tabular*}{\hsize}{l l l}
    \hline
	& Forward & Backward\\
	\hline
	Base & $F_0 = I$  &  $B_0 = \neg P$\\
	Induction & $F_{i+1} = R(F_i)$  & $B_{i+1} = R^{-1}(B_i)$\\
	Safe Check & $F_{i+1} \subseteq \bigcup_{0\leq j\leq i} F_j$  &  $B_{i+1} \subseteq \bigcup_{0\leq j\leq i} B_j$\\
	Unsafe Check & $F_i\cap \neg P \not=\emptyset$  &  $B_i\cap I \not= \emptyset$ \\
	\hline
\end{tabular*}
\end{table}

For forward search, $F_i$ denotes the set of states that are reachable from $I$ within $i$ steps, which is computed by iteratively applying $R$. At each iteration, we first compute a new $F_i$, and then perform safe checking and unsafe checking. If the condition in the safe/unsafe checking is satisfied, the search process terminates.
Intuitively, unsafe checking $F_i\cap \neg P \neq \emptyset$ indicates that some bad states are within $F_i$ and safe checking $F_{i+1}\subseteq \bigcup_{0\leq j\leq i} F_j$ indicates that all the reachable states from $I$ have been checked and none of them violate $P$.
For backward search, the set $B_i$ is the set of states that can reach $\neg P$ in $i$ steps, and the search procedure is analogous to the forward one.

\subsection{\SAT Solving and Unsatisfiable Cores}

In propositional logic, a \emph{literal} is an atomic variable or its negation. A \emph{cube} (resp. \emph{clause}) is a conjunction (resp. disjunction) of literals. The negation of a clause is a cube and vice versa. A formula in \textit{Conjunctive Normal Form} (CNF) is a conjunction of clauses. For simplicity, we also treat a CNF formula $\phi$ as a set of clauses. Similarly, a cube or a clause $c$ can be treated as a set of literals or a Boolean formula, depending on the context.
    
We say a CNF formula $\phi$ is satisfiable if there exists an assignment of each Boolean variable in $\phi$ such that $\phi$ is true; otherwise, $\phi$ is unsatisfiable. A \SAT solver can decide whether a CNF formula $\phi$ is satisfiable or not. It emits a Boolean assignment to the variables, called a model of $\phi$, if $\phi$ is satisfiable. Otherwise, it emits an unsatisfiable core as explained in the introduction, based on a subset of the assumptions.

\subsection{Complementary Approximate Reachability (\CAR)}\label{sec:CAR}

\CAR is a relatively new \SAT-based safety model checking approach that is essentially a reachability-analysis algorithm, inspired by \PDR~\cite{LZZPV17}. Unlike \BMC~\cite{BCCFZ99,BCCZ99}, \CAR is complete, i.e., it can also prove correctness. \CAR maintains two sequences of state sets (also called `frames'), that are defined as follows:

\begin{definition}[Over/Under Approximating State Sequences]\label{def:OU}
Given a transition system $Sys=(V,I,T)$ and a safety property $P$, the over-approximating state sequence  $O \equiv O_0, O_1,\dots, O_i~(i \geq 0)$, and the under-approximating state sequence $U \equiv U_0, U_1,\dots, U_j~(j \geq 0)$ are finite sequences of state sets such that, for $k\geq 0$:

\begin{table}[!htb]
\renewcommand{\arraystretch}{1.4}
\label{tab:O/U Sequence}
\centering
\begin{tabular*}{\hsize}{l l l}
    \hline
    & ~~~~$O$-sequence & ~~~~$U$-sequence \\
    \hline

    Base: & ~~~~$O_0 = \neg P$ &   ~~~~$U_0 = I$\\
    Induction: & ~~~~$O_{k+1} \supseteq R^{-1}(O_{k})$ & ~~~~$U_{k+1} \subseteq R(U_{k})$ \\
    Constraint: & ~~~~$O_k\cap I=\emptyset$ &~~~~ $--$\\

    \hline
\end{tabular*}
\end{table}
\noindent 
These sequences determine the termination of \CAR as follows: 

\begin{itemize}
    \item Return `Unsafe' if $\exists i\cdot U_i\cap \neg P\not = \emptyset$.
\item Return `Safe' if $\exists i\geq 1 \cdot (\bigcup_{j=0}^i O_j) \supseteq O_{i+1}$. 
\end{itemize}
\end{definition}

Notably, \CAR can also use the over and under approximating sequences reversed, i.e., use the over-approximating sequence in the forward direction, from the initial state towards the negated property, while using the under-approximating sequence from the negated property towards the initial state. In this paper, we only consider the direction as stated in Definition~\ref{def:OU} (this was called `backward \CAR' in~\cite{LZZPV17,LDPRV18}). 

At the high level, \CAR can be considered a general version of \PDR, as the O-sequence in \CAR is not necessarily monotone, while that in \PDR is. As a result, \CAR can have a more flexible methodology for the \emph{state generalization}, i.e., directly using the \UC from the \SAT solver rather than computing the \emph{relative inductive clauses}. However, \CAR needs to invoke additional \SAT queries to find the invariant (checking safety), while  \PDR can do it with a simple syntactic check.

\begin{algorithm}
\caption{Complementary Approximate Reachability (\CAR). }\label{alg:car}
\LinesNumbered
\DontPrintSemicolon
\KwIn{A transition system $Sys=(V, I, T)$ and a safety property $P$}
\KwOut{`Safe' or (`Unsafe' + a counterexample)}

\lIf*{$SAT(I\wedge \neg P)$}
{
    \textbf{return} `Unsafe'\;
}
$U_0\coloneqq I$, $O_0\coloneqq\neg P$\; \label{alg:car:init}
\While{true \label{alg:car:outerloop}}{

 $O_{tmp} \coloneqq \neg I$  \label{alg:otmp}
 
\While{$state\coloneqq pickState(U)$ is successful}
{\label{alg:car:mainloopbegin}\label{alg:car:pickstate}
    
    $stack \coloneqq \emptyset$\;
    
    $stack.push(state,|O|-1)$\;    
    
    \While{$|stack|\neq 0$}
    {
        
         $(s,l)\coloneqq stack.top()$\tcp*[f]{Assume~$s \in U_j$}
         
        \lIf*{$l < 0$}
            {
                \Return `Unsafe'\;\label{alg:car:unsafe}
            }
        \textcolor{red}{$\hat{s}$ = Reorder (s, l + 1)}\;\label{alg:car:reorder}
        \If{$SAT(\hat{s}, T \wedge  O_l')$  \label{alg:car:satcall}
        }
        {
            $t\coloneqq GetModel()|_{V'}$\label{alg:car:satcallyesbegin}
            
            $U_{j+1}\coloneqq U_{j+1}\cup t$ \tcp*[f]{Widening $U$}
            
            $stack.push(t,l-1)$\;
        \label{alg:car:satcallyesend}}  
        \Else
        {
            $stack.pop()$\;\label{alg:car:satcallnobegin}
            $uc\coloneqq getUC()$\;
            \lIf* {$l+1 < |O|$} 
        {$O_{l+1}\coloneqq  O_{l+1}\wedge(\neg uc)$ \label{alg:car:satcallnoend}} 
        
        \lElse {$O_{tmp}\coloneqq  O_{tmp}\wedge(\neg uc)$} 
            
            \lWhile {$l+1 < |O|$ and $\neg s \in O_{l+1}$ }
            {
                $l := l+1$  \label{alg:car:skip}
            }
            
            \lIf*{$l+1 < |O|$} {\label{alg:car:rollbackbegin} $stack.push(s, l)$   }\label{alg:car:rollbackend} 
        }
    
    }
    }
    \lIf*{$\exists i\geq 1$ s.t. $(\bigcup_{0\leq j\leq i} O_j) \supseteq O_{i+1}$ \label{alg:car:safe}}
    {\Return `Safe'\;} 
    
    Add a new state-set to $O$ and initialize it to $O_{tmp}$\;
}\label{alg:car:mainloopend}

\end{algorithm}

Algorithm \ref{alg:car} describes \CAR. It progresses by widening the $U$ sets, and narrowing the $O$ sets, which are initialised at Line~\ref{alg:car:init} to $I$ and $\neg P$, respectively. The algorithm maintains a stack of pairs $\langle state, level\rangle$ where $level$ refers to an index of an $O$ frame.
$O_{tmp}$, initialised to $\neg I$ in Line~\ref{alg:otmp} and later updated, represents the next frame to be created. 

Initially, a state from the $U$-sequence is heuristically picked (Line \ref{alg:car:pickstate}) -- by default from the end to the beginning --  and pushed to the stack. In each iteration of the internal loop,  
\CAR checks whether the state at the top of the stack, call it $s$, can transition to the $O_l$ frame. This is done by checking if $s\wedge T\wedge {O_l}'$ is satisfiable (Line~\ref{alg:car:satcall}, $\hat{s}$ is exactly $s$ if literal-ordering is not invoked, otherwise a reordered version). If yes, a new state $t\in O_l$ is extracted from the model to update the $U$-sequence (Line~\ref{alg:car:satcallyesbegin}-\ref{alg:car:satcallyesend}), effectively \emph{widening} it; Otherwise, the negation of the unsatisfiable core is used to constrain the $O$ frame of $s$ (level $l+1$), effectively \emph{narrowing} it (Lines~\ref{alg:car:satcallnobegin}-\ref{alg:car:satcallnoend}), and pushing $s$ back to the stack. In Line~\ref{alg:car:skip}, \CAR skips frames that already block $s$. 

\CAR returns `Unsafe' as soon as the working level $l$ is less than 0, which indicates that a bad state in $\neg P$ is reached (line~\ref{alg:car:unsafe}). Otherwise, \CAR returns `Safe' if the $O$ sequence includes all the states that can reach $\neg P$ -- this is checked via the condition in Line~\ref{alg:car:safe}, which was also mentioned as part of Definition~\ref{def:OU}. 

%% file: 3.revisit_prior_work.tex
\section{Literal Reordering Strategies: prior work and insights}\label{sec:revisit}

\begin{example}[Prior literals are more likely to appear in the \UC]\label{exp:one}
Let the clauses and assumptions be 
\begin{eqnarray*}
\phi & \doteq &\{ (a_1\vee \neg a_4 \vee \neg a_5),  (a_3\vee \neg a_4 \vee \neg a_5 ), (a_2 \vee a_4) \} \\
\mathcal{A} & \doteq & (\neg a_1, a_2,\neg a_3, a_4, a_5)\;.
\end{eqnarray*} 
Different literal orderings generate different \UCs:
\label{eg:literal_ordering_matters}
\begin{align*}
\textbf{\emph{Order 1:}} \  Assum_1=& \ (\neg a_1, a_2, a_4, a_5, \neg a_3) \\
BCP(\neg a_1):& \{ ( False \vee \neg a_4 \vee \neg a_5 ),\\& \medspace (a_3\vee \neg a_4 \vee \neg a_5 ),(a_2 \vee a_4) \}\\
BCP(a_2):&\{ ( False \vee \neg a_4 \vee \neg a_5 ),\\& \medspace (a_3\vee \neg a_4 \vee \neg a_5 ),(True \vee a_4) \}\\
BCP(a_4):&\{ ( False \vee False \vee \neg a_5 ),\\& \medspace (a_3\vee False \vee \neg a_5 ),(True \vee a_4) \}\\
BCP(a_5):&\{ ( False \vee False \vee False ),\\& \medspace (a_3\vee False \vee False ),(True \vee a_4) \}\\
 \longrightarrow \UC_1 = &\ ({\neg a_{1}}, {a_{4}}, {a_{5}}). \\
\textbf{\emph{Order 2:}} \ Assum_2=& \ (a_{5}, a_{4}, \neg a_{3}, a_2, \neg a_1)\\
BCP(a_5):&\{ (a_1\vee \neg a_4 \vee False ), \\&(a_3\vee \neg a_4 \vee False ),(a_2 \vee a_4) \}\\
BCP(a_4):&\{ (a_1\vee False \vee False ), \\&(a_3\vee False \vee False ),(a_2 \vee a_4) \}\\
BCP(\neg a_3):&\{ (a_1\vee False \vee False ),\\&(False\vee False \vee False ),(a_2 \vee a_4) \}\\
\quad  \longrightarrow \UC_2 = & \ ({a_{5}}, {a_{4}}, {\neg a_{3}}).
\end{align*}
\end{example}

As mentioned in the introduction, modern CDCL-based \SAT solvers such as the derivatives of \tool{Minisat}~\cite{ES03} take as input, in addition to the formula $\phi$, a vector of literals $\mathcal{A}$, called the assumptions, and checks whether $\bigwedge_{l\in \mathcal{A}}l\wedge \phi$ is satisfiable. 

The \SAT solver chooses the assumption literals to be the first decisions in the order they are given. As usual, after each such decision, it invokes BCP.
Suppose there is already a conflict in the first $|A|$ ($A\subseteq \mathcal{A}$) decision levels (recall that this can happen after learning and backtracking to those levels). In that case, the search is terminated -- the formula is declared unsatisfiable under $A$. The solver can be asked to analyse the cause of the conflict and return it in the form of a subset of $A$.
This implies that assumption literals after $|A|$ in the predefined order cannot be part of the generated \UC.
As a result, prior assumption literals have a higher probability of appearing in the \UC. That is why literal ordering matters. 
Example \ref{exp:one} illustrates this point.


\textbf{The \textbf{Intersection} strategy and \textbf{locality} of cores:} 
The \emph{Intersection} strategy~\cite{DLPVR19} places the intersection with the last \UC in the beginning of the assumptions sequence -- see Algorithm \ref{alg:intersection}. It is an implementation of the \emph{Reorder} 
function that is called in line~\ref{alg:car:reorder} of Algorithm~\ref{alg:car} with level $l+1$, namely the previous level. 
In line~\ref{alg:intersection:icube} of Algorithm~\ref{alg:intersection}, the literals from this \UC are placed first in the order, which makes them more likely to appear in the new core, hence make consecutive cores similar. This is what we call `the \emph{locality} of the cores'. 

The term \emph{locality} is used, among other places, in describing decision heuristics in SAT solving. All CDCL solvers use decision heuristics that prioritize variables that participated in recent conflicts, hence they focus the search. Although this is not directly related to the current paper, our hypothesis is that this decision strategy is effective because it generates proofs faster: similar clauses are necessary for constructing a resolution proof (for satisfiable cases, learning has little effect to begin with~\cite{O15}). And if there is a small core, it is better to focus the search and hopefully find it rather than generating unrelated clauses.  

Our argument is that finding cores in \CAR that are similar should have a similar effect: it makes proofs involving the $O$  frames easier and hence faster. In other words, every time that we check whether a state can reach an $O$ frame, if that frame contains apriori many of the clauses that are necessary for the proof that the state is not reachable, the proof will converge faster. We tested this hypothesis empirically, and the results appear in Table~\ref{tab:speed}. While the first row shows the effect of locality on the average run time of UNSAT cases, the second row is the average overall time for proving that a state cannot reach an $O$ frame, i.e., the average time of an iteration of the loop in line~\ref{alg:car:pickstate} of Algorithm~\ref{alg:car}. 
The evaluation is based on benchmarks from the single safety property track of the 2015~\cite{hwmcc15} and 2017~\cite{hwmcc17} Hardware Model Checking Competition (HWMCC~\cite{8102233})\footnote{The experiment setup in this section is the same as that in Sec.~\ref{sec:experiments}.}.

\begin{table}
    \centering
    \renewcommand{\arraystretch}{1.2}
    \caption{The Intersection strategy, through what we call locality, accelerates UNSAT calls and proof finding \label{tab:speed}}
    \begin{tabular}{c|c|c} \tline
        Strategy        & Natural  & Intersection  \\ \tline
        Average time of UNSAT calls(s)            & 0.0132                          & 0.0105                       \\ \hline
        Average time of finding proofs(s)    & 0.9541                          & 0.6287	                     \\ \tline
        
\end{tabular}
\end{table}

\begin{algorithm}\label{alg:intersection}
\caption{Reordering algorithm: \emph{Intersection}}
\LinesNumbered
\DontPrintSemicolon
\KwIn{A vector of literals $s$ representing a state, and the frame level $l$}
\KwOut{$\hat{s}$: the reordered $s$}
$\hat{s} = \emptyset$;

$lastUC \coloneqq getLastUC(l)$; \Comment{A vector of literals} \label{alg:intersection:lastuc}

\For{each $lit \in lastUC$}
{
    \If{ $lit \in s $}
    {
        $\hat{s}.pushBack(lit)$ \;\label{alg:intersection:icube}
    }
}

\For{each $lit \in s \wedge lit \notin \hat{s}$ } 
{
    $\hat{s}.pushBack (lit)$\;
}
\Return $\hat{s}$
\end{algorithm}

\textbf{The \textbf{Rotation} strategy:}
The \emph{Rotation} technique — demonstrated in Algorithm \ref{alg:rotation} — maintains a vector $common$ for each level $l$ to track the similarity among recent failed states, i.e., the $common$ vector is a reordered version of the last failed state, with the intersection of the failed states in the front. 

The key insight behind \emph{Rotation}~\cite{DLPVR19} is that in cases where the solver consistently returns similar states that share common literals but fail to explore deeper levels, the search process may be trapped within a specific sub-space. Consequently, Rotation prioritises the common part in the front, intending to generate a \UC from it, thus facilitating an exit from the problematic sub-space.

\begin{algorithm}\label{alg:rotation}
\caption{Reordering algorithm: Rotation}
\LinesNumbered
\DontPrintSemicolon
\KwIn{A vector of literals $s$ representing a state, and the frame level $l$}
\KwOut{$\hat{s}$: the reordered $s$}
$\hat{s} = \emptyset$;

$cVec \coloneqq getCommonVector(l)$; \Comment{get common vector}

\For{each $lit \in cVec$}
{
    \If{ $lit \in s $}   
    {
        $\hat{s}.pushBack(lit)$\;\label{alg:rotation:rcube}
    }

}

\For{each $lit \in s \wedge lit \notin \hat{s}$ } 
{
    $\hat{s}.pushBack(lit)$ \;
}

\Return $\hat{s}$

// Future updates:

\If{$ \neg SAT(\hat{s}, T \wedge  O_l')$ \Comment{Fail to reach}\;}
{
    $setCommonVector(l,\hat{s})$ \Comment{Update common vector}
}

\end{algorithm}

\begin{figure*}
\centering
{\includegraphics[
width=0.8\textwidth, height=6.5cm
]{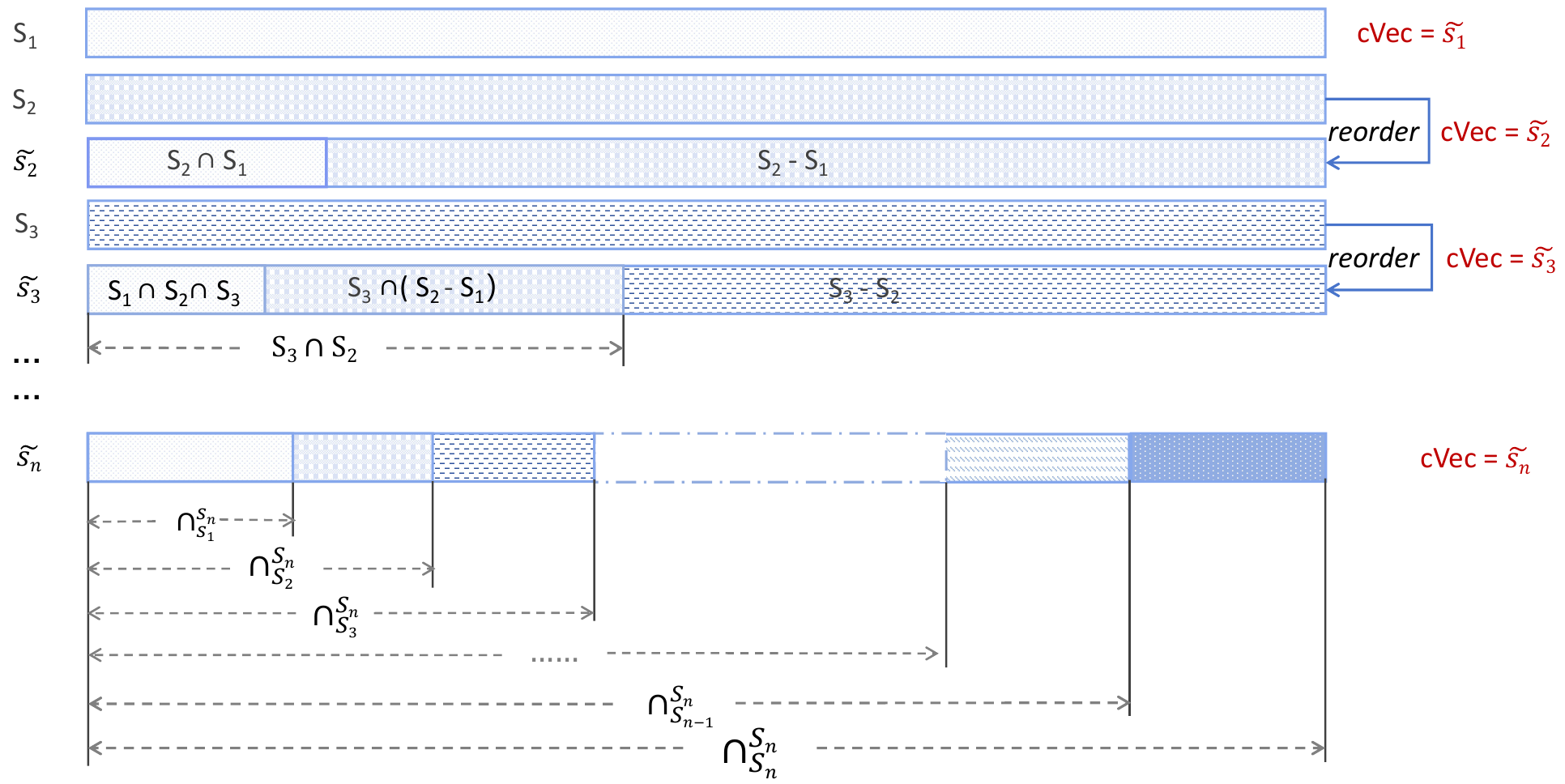}}
\caption {The updating process of $cVec$, which is the core of \emph{Rotation}. The consecutive intersection of recent failed states ($\bigcap{s_i}$) is in the very beginning. }
\label{fig:RotateIllustration}
\end{figure*}

While maintaining the consecutive intersection of all the recent failed states is feasible, the length of the joint part decreases, consequently containing diminishing information. As a remedy, it was suggested in~\cite{DLPVR19} to preserve a fixed length of the $common$ vector. As demonstrated in Fig. \ref{fig:RotateIllustration}, $common$ always preserves the last failed state -- and is reordered to encode the historical information. While the first segment can be regarded as the intersection of all states ($\bigcap^{s_n}_{s_1}$, as shown in the figure), the combination of the first two segments can be seen as all the states except one ($\bigcap^{s_n}_{s_2}$), and so forth. 

The rational behind \emph{Rotation}, is that it encourages finding \UCs made of literals that appeared in many recent failed states, hence it diverts the search from areas that seem to lead nowhere. This, in turn, should reduce the number of states that are checked and the corresponding number of SAT calls. Indeed, our results in Table~\ref{tab:rotation:speed} show a reduction in the number of SAT calls with \emph{Rotation} when proving that a state from a $U$ frame cannot reach any of the $O$ frames (i.e., a single iteration of the loop starting in line~\ref{alg:car:mainloopbegin} of Algorithm~\ref{alg:car}). 
The basis of the evaluation is the same as in Table~\ref{tab:speed}. 

\begin{table}
    \centering
    \renewcommand{\arraystretch}{1.2}
    \caption{The Rotation strategy helps escaping bad areas of the search, and consequently it lowers the number of SAT queries and the time of proofs.
\label{tab:rotation:speed}}
    \begin{tabular}{c|c|c} \tline
        Strategy:                                & Natural   & Rotation  \\ \tline
        Average \#SAT calls to find proofs    & 207.13    & 190.25    \\ \hline
        Average time to find proofs (s)       & 0.9541    & 0.7277    \\ \tline
\end{tabular}
\end{table}

\begin{figure}
\centering
{\includegraphics[width=0.48\textwidth]{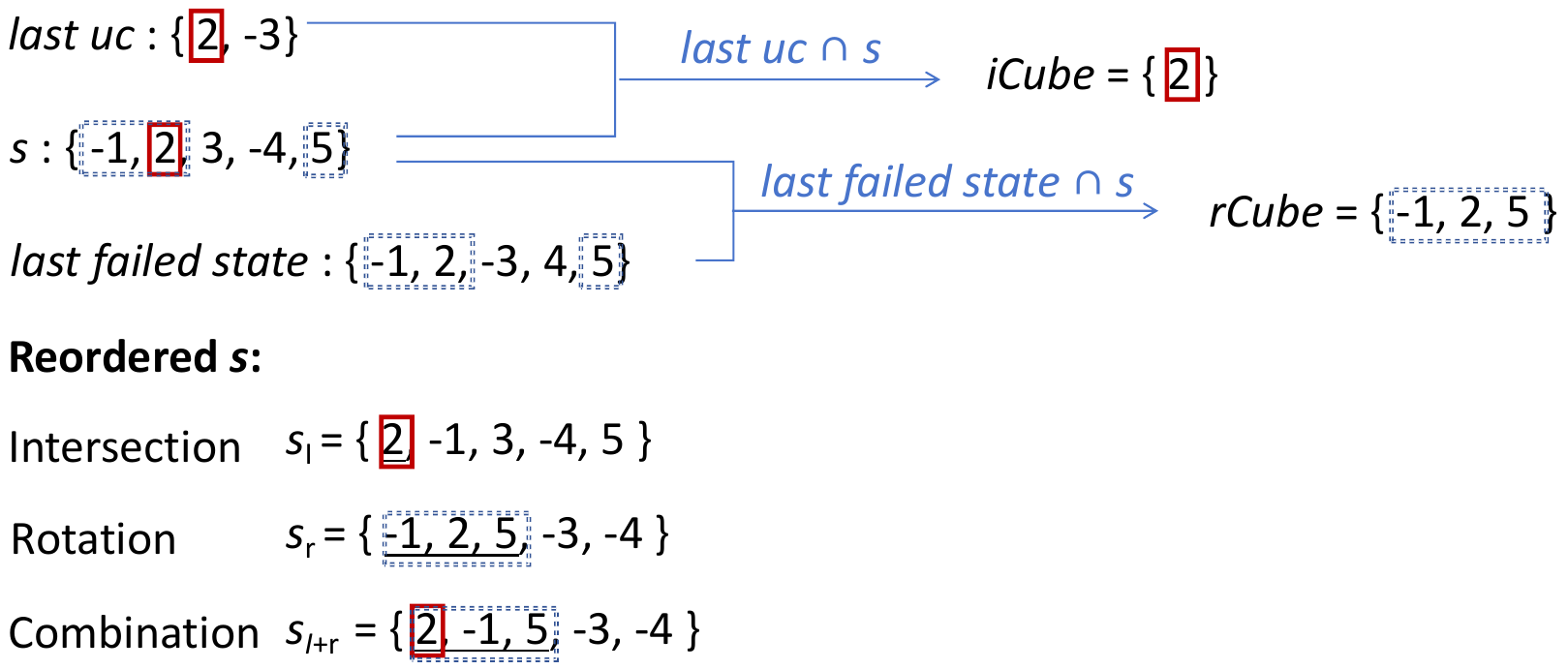}}
\caption {An example of the reordering process, where $s_{i}$, $s_{r}$ and $s_{i+r}$ each represents the reordered state $\hat{s}$ using only \emph{Intersection}, only \emph{Rotation} and their combination,  respectively.}
\label{fig:ReorderExample}
\end{figure}

\textbf{Combining \emph{Intersection} and \emph{Rotation}:}
When it comes to combining these two algorithms, it should be noted that the latest \UC selected in \emph{Intersection} is derived from the last failed state. This observation leads to the conclusion that the $iCube$ (the cube generated via \emph{Intersection}) is a subset of the $rCube$ (the cube generated via \emph{Rotation}). 
As shown in Fig.\ref{fig:ReorderExample}, to integrate the two algorithms is merely to position the literals produced by \emph{Intersection} ahead of those generated by \emph{Rotation} while  eliminating duplicate literals in the latter.

\begin{table}
    \centering
    \renewcommand{\arraystretch}{1.2}
    \caption{The Combination of Intersection and Rotation can find proofs faster. \label{tab:combine} }
    \begin{tabular}{c|c|c} \tline
        Strategy:                               & Natural   & Combination (I+R)  \\ \tline
        Average time of UNSAT calls(s)          & 0.0132    & 0.0137            \\ \hline
        Average \#(SAT Query) to find proof     & 207.13    & 173.57            \\ \hline
        Average time of finding proofs(s)       & 0.9541    & 0.5990            \\ \tline
\end{tabular}
\end{table}

The results in Table \ref{tab:combine} show that indeed the combination finds proofs faster on average.

%% file: 4.new_approaches.tex
\section{Literal Ordering Strategies: new approaches}\label{sec:newapproach}

\input{4.1Local}

\input{4.2HybridCAR}

%% file: 4.1Local.tex
As discussed in the previous section, \emph{Intersection} and \emph{Rotation}, either separately or combined, determine the position of only a limited portion of the whole literal set, whereas the position of other literals is determined by what we called the `natural' order, which is arbitrary.

A motivating example is depicted in Fig.\ref{fig:RotateIllustration}. As shown, the segment $S_3-S_2$ within $\hat{S_3}$ remains unaffected by both literal ordering strategies, i.e., it just remains the natural  order, yet it constitutes more than half of the state's length. 

In this section we will show a way to increase the portion of literals that their position is determined, utilising more historical information on the cores, and consequently improving the overall runtime. 

\subsection{Literal reordering with \local}
\label{subsection:our new reorder}
Stemming from the intuition that incorporating recent \UCs beyond the most recent one could help by improving locality, we propose a new literal ordering strategy \emph{\local}, which is outlined in Algorithm \ref{alg:promoted_reorder}. By expanding the scope of considered \UCs, intersecting with each and organizing the results chronologically (with the intersection with newer \UCs placed earlier), \emph{\local} facilitates sorting a greater number of literals, thereby refining the guidance of the search.


As shown in the algorithm, in addition to the state $s$ and frame level $l$, a new parameter $iLimit$ is introduced to denote the limit on the amount of \UCs to utilise. The \emph{for} block at Line \ref{alg:newreorder:icube:begin}-\ref{alg:newreorder:icube:end} computes the intersection according to the corresponding \UC, and pushes them into $\hat{s}$ in order. For the \emph{if} block at Line \ref{alg:newreorder:rcube:mid}-\ref{alg:newreorder:rcube:end}, it is similar to \emph{Rotation}  .

\begin{algorithm}
\caption{Reordering algorithm: \emph{\local} }\label{alg:promoted_reorder}
\LinesNumbered
\DontPrintSemicolon
\KwIn{A state $s$, frame level $l$, configuration $iLimit$ }
\KwOut{$\hat{s}$: The reordered $s$}

$\hat{s} \coloneqq \emptyset$ \\
\For{$k: 0 \rightarrow iLimit$ \label{alg:newreorder:icube:begin}}
{
    Let $ref_k = getTheLast\_kth\_UC(l)$\label{alg:newreorder:getlastUC} \; 
    \If{$ref_k \neq \emptyset$}
    {
    \For{each $lit \in ref_k$}
    {
        \If{ $lit \in s \wedge lit  \notin \hat{s} $}   
        {
            $\hat{s}.pushBack(lit)$\;\label{alg:reorder_i} \Comment{Literals added here form the $iCube$s}
        }
    }
    }
}\label{alg:newreorder:icube:end}

$cVec \coloneqq getCommonVector(l)$\; 

\For{each $lit \in cVec$\label{alg:newreorder:rcube:begin}}
{
    \If{ $lit \in s \wedge lit \notin \hat{s} $ \label{alg:newreorder:rcube:mid}}  
    {
        $\hat{s}.pushBack(lit)$\;\label{alg:newreorder:rcube} \Comment{Literals added here form the $rCube$}
    }

}\label{alg:newreorder:rcube:end}

\For{each $l \in s \wedge l \notin \hat{s}$} 
{
    $\hat{s}.pushBack (l)$\;
}

\Return $\hat{s}$
\end{algorithm}

\begin{figure}
\centering
{\includegraphics[width=0.45\textwidth]{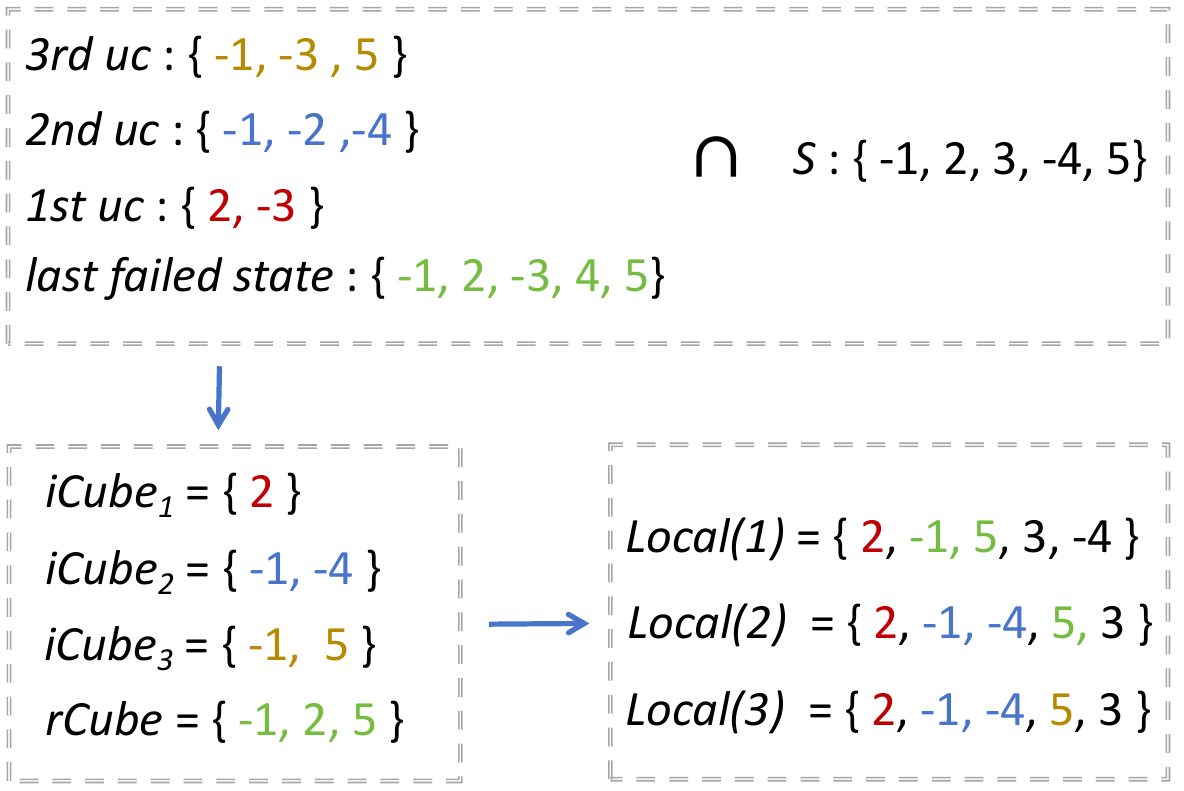}}
\caption {An example of the \emph{\local} strategy. $Local(k)$ denotes reordered state utilizing $k$ \UCs, i.e., $iLimit = k$.
\label{fig:LocationExample}}
\end{figure}

\begin{example}
Fig.~\ref{fig:LocationExample} illustrates a computational process for the \emph{\local} strategy with several different $iLimit$ values. The upper dashed box in the figure shows the last $3$ \UCs in chronological order (1st being the most recent one), along with the last failed state, and the current state $s$. Next, $iCubes$ and $rCube$ are computed based on the above data, similar to the calculation in Fig.~\ref{fig:ReorderExample}, as shown in the lower left dashed box. Finally, in the lower right dashed box, $s$ is reordered by $iCubes$ and $rCube$ based on different choices of $iLimit$. As is shown, by incorporating the 2nd \UC, the literal `-4' is successfully impacted. 
\qed
\end{example}

\begin{figure}
\centering
{\includegraphics[width=0.48\textwidth]{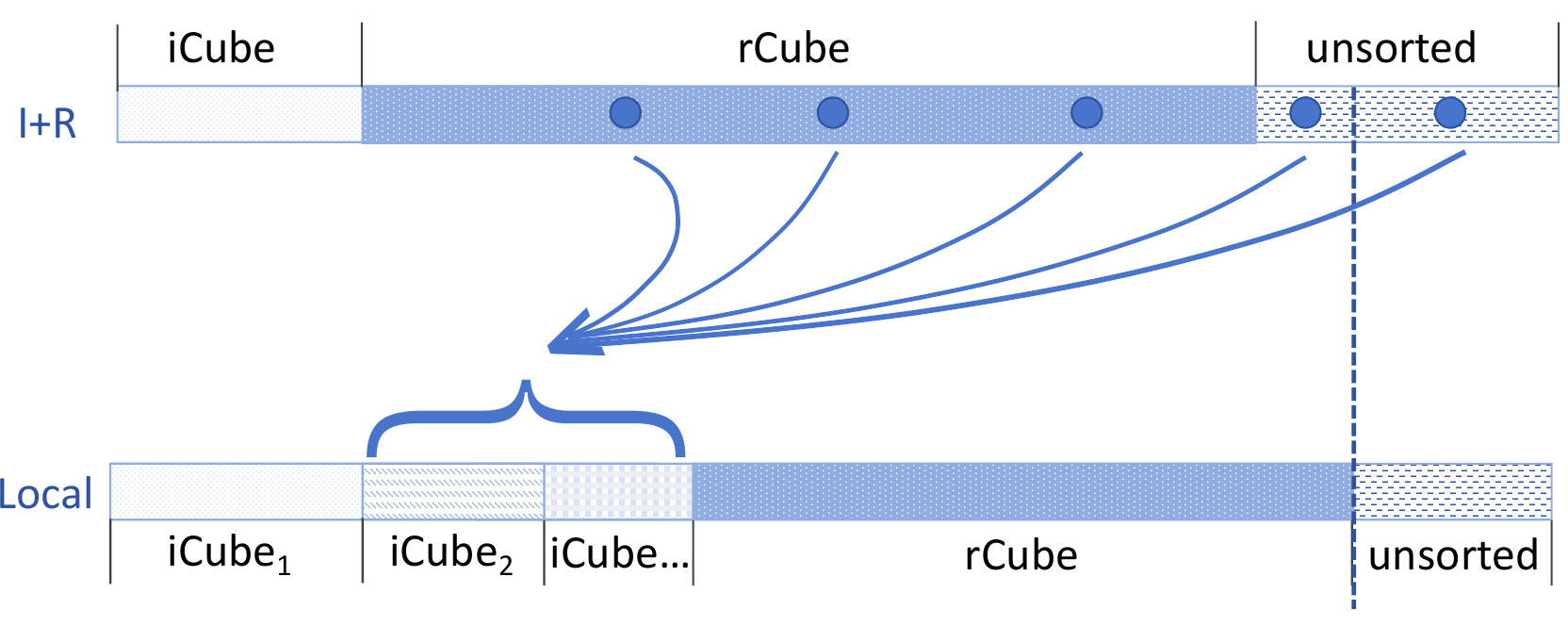}}
\caption {The difference between \emph{\local} and combination of \emph{Intersection} and \emph{Rotation}. Some literals (blue dots) from $rCube$ and $unsorted$ are prioritised.}
\label{fig:LocalIllustration}
\end{figure}

The distinction between \emph{\local} and \emph{Intersection} is demonstrated in Fig.~\ref{fig:LocalIllustration}). It prioritises literals that would otherwise be relegated to the rear of $rCube$, or even after $rCube$.

\begin{figure}
\centering
{\includegraphics[width=0.48\textwidth]{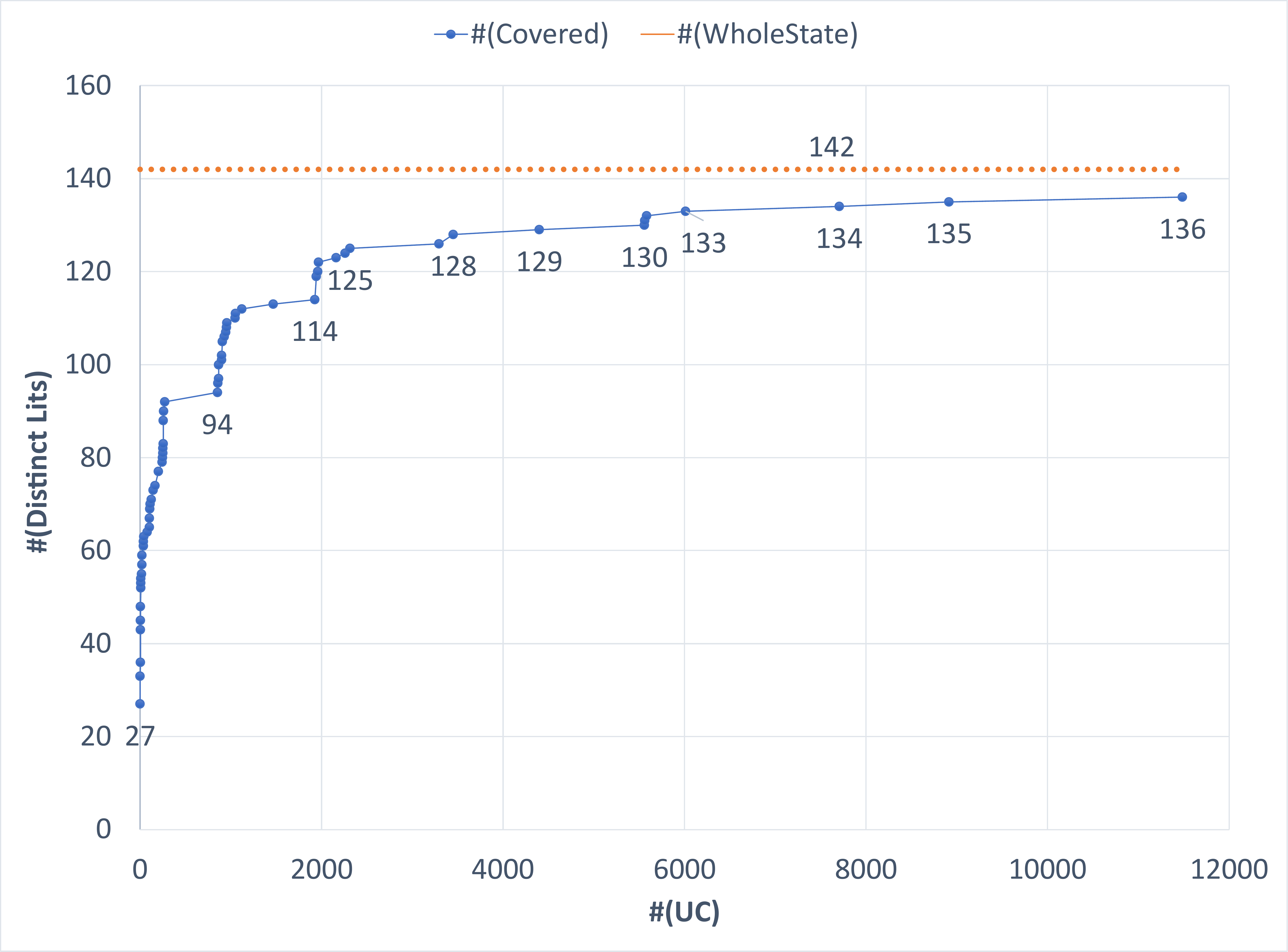}}
\caption {Increasing the number of \UCs that are intersected has a diminishing effect on the number of distinct literals that are covered (this figure was created based on a formula from 6s33.aig).}
\label{fig:UC&Lits}
\end{figure}

\textbf{Tuning \emph{\local}:} 
In the \emph{\local} strategy, the parameter $iLimit$, which denotes the maximum number of utilised \UCs, serves as a metric of the `local' scope, defining the range within which a \UC is considered `recent'. In other words, given that the relevance of a \UC to the current query diminishes as it becomes more distant, setting a limit excludes prior outdated \UCs from current consideration. While increasing the value of $iLimit$ allows for the inclusion of additional information, it also diminishes the impact of \emph{Rotation} due to the precedence of $iCubes$ over the $rCube$. Furthermore, 
while it is feasible to set the $iLimit$ large enough to order all the literals, this approach is observed to be highly inefficient. The considerable increase in cost to get one more literal reordered, i.e., one that appears in a subsequent \UC, but not in any previous one, often necessitates  thousands or even tens of thousands of \UCs. This phenomenon is demonstrated in Fig.\ref{fig:UC&Lits} for a particular formula.

The optimal value of $iLimit$ depends, of course, on the specific problem context and constitutes a trade-off between literal coverage and the computational cost to achieve it. Indeed, the results in Table \ref{tab:local:speed} (middle column) show that the speed to find proofs of \emph{\local} depends on $iLimit$ but as expected, it is not monotonic. They also show that with these low values of $iLimit$ we are able to find proofs faster than the previous methods. 
In practice the best $iLimit$ value can be found based on experiments, but there is also an option to change it during run time, as we will explain later.


\begin{table}
    \centering
    \renewcommand{\arraystretch}{1.2}
    \caption{\local can find proofs even faster. `With' and `Without' refers to the `conflict literal' optimization. \label{tab:local:speed} }
    \setlength{\tabcolsep}{0.6cm}
    \begin{tabular}{c|c|c} \tline
     \multirow{2}{*}{Strategy}
          & \multicolumn{2}{c}{Average time of finding proofs(s)}      \\ \cline{2-3}
                            & Without   & With                         \\ \tline
        Natural             & 0.9478    & 0.9541                       \\ \hline
        Combination(I+R)    & 0.7728    & 0.5990                       \\ \hline
        Local(2)            & 0.6191    & 0.5887                       \\ \hline
        Local(3)            & 0.6476    & 0.5566                       \\ \hline
        Local(4)            & 0.5721    & 0.4836                       \\ \hline
        Local(5)            & 0.6413    & 0.5078                       \\ \tline
\end{tabular}
\end{table}

\subsection{Moving forward the conflict literals}\label{sec:conflictliterals}
Not all literals in a \UC are equal. Specifically, the last literal added to the core, by definition, was \emph{necessary} for that proof (in other words, it is part of a \emph{minimal} core). We call it the \emph{conflict} literal. During the process of getting \UCs, we give such conflict literals a higher priority by placing them in the front of the core. For example, suppose that $iCube = (1,2,3)$ and literal 3 is the conflict literal of this clause. Then we re-order $iCube$ to
$iCube = (3,1,2)$, before we proceed with building the assumption literal order as described earlier (Algorithm~\ref{alg:promoted_reorder}).
As shown in the right column of Table.~\ref{tab:local:speed}, this small optimization accelerates, on average, the process of finding proofs. 

From here on, when we say \emph{\local}, we mean \emph{\local} together with this optimization.

%% file: 4.2HybridCAR.tex
\subsection{\approach: Combining different orderings}\label{sec:approach}
Empirically, the best configuration of \emph{\local} varies according to the specific problem and is hard to predict. It is often observed that a model checking problem that can be solved easily with one literal ordering strategy will time-out with another. This encouraged us to research a dynamic strategy, by which the configuration is periodically switched.

\begin{table}[ht]
    \centering
    \renewcommand{\arraystretch}{1.4}
    \caption{The U-sequence expands quickly and slows down the iteration time in 6s33.aig \label{tab:caseU}}
    \begin{tabular}{c|c|c} \tline
        Round & TimeForThisRound(s) & size(U) \\ \tline
        1 & 0 & 1 \\ \hline
        2 & 0 & 2 \\ \hline
        3 & 0.011 & 15 \\ \hline
        4 & 0.007 & 20\\ \hline
        5 & 0.763 & 823\\ \hline
        6 & 7.392 & 5625\\ \hline
        7 & 12.264 & 9632\\ \hline
        8 & 43.158 & 19285\\ \hline
        9 & 356.92 & 57263\\ \hline
        10 & $>$3000(Timeout)& 110235 \\ \tline
        
    \end{tabular}
    
\end{table}

We coupled this direction with a new restart mechanism for \CAR. The U-sequence in \CAR is observed to expand quickly, resulting in increasingly longer time to extend a new $O$ frame.

Taking a closer look at the particular case shown in Table \ref{tab:caseU}, it can finish the first five rounds within 1 second when the size of the U-sequence is moderate, but then gets stuck in the 10th round, where it spends more than 50 minutes. Indeed, all the states in the $U$ frame are explored (see line \ref{alg:car:mainloopbegin} in Algorithm~\ref{alg:car}) before a new $O$ frame can be opened in line~\ref{alg:car:mainloopend}. Perhaps resetting the $U$ frame and progressing with a different literal ordering can converge faster. 

\begin{algorithm}[htbp]
\caption{\approach. }\label{alg:hybridcar}
\LinesNumbered
\DontPrintSemicolon
\KwIn{A transition system $Sys=(V, I, T)$ and a safety property $P$, time limit to restart $TimeLimit$}
\KwOut{`Safe' or (`Unsafe' + a counterexample)}

\lIf*{$SAT(I\wedge \neg P)$}
{
    \textbf{return} `Unsafe'\;
}
$U_0\coloneqq I$, $O_0\coloneqq\neg P$; 

\While{true}{

 $O_{tmp} \coloneqq \neg I$ 
 
\While{$state\coloneqq pickState(U)$ is successful \label{alg:pickState} }
{
    
    $stack \coloneqq \emptyset$\;
    
    $stack.push(state,|O|-1)$\;    
    
    \While{$|stack|\neq 0$}
    {
        
         $(s,l)\coloneqq stack.top()$

        \begin{elaboration}
                \If{$timeExceed(TimeLimit)$}
                    {
                        $conf \coloneqq getNextConfig()$
                        \\
                        $restart()$;\label{alg:restart}

                    }
        \end{elaboration}

        \lIf*{$l < 0$}
            {
                \Return `Unsafe'\;
            }
        \textcolor{red}{$\hat{s}$ = Reorder (s, l + 1, conf)}\;\label{alg:hybridcar:reorder}
        \If{$SAT(\hat{s}, T \wedge  O_l')$  } 
        {
        ... \Comment{Same as in original CAR}
        }

    }
}
    ...
    
}

\end{algorithm}

Algorithm \ref{alg:hybridcar} shows our implementation of \approach, based on the observations above. The only difference between \approach and \CAR falls in the dotted rectangle. A timer is kept in the SAT solver, starting at the beginning, to calculate the time consumption. Once it exceeds the time limit, it triggers the restart procedure. In the \emph{restart} procedure, it only keeps the states in the lowest level in the $U$-sequence, clears all the others, and switches the reordering configuration. After that, it resets the timer, jumps out of the current loop, and starts searching with a new configuration from the beginning. 
In the \emph{getNextConfig} procedure, we employ a simple strategy to change the demarcation of locality: increase the number of $iLimit$ by one.
Finally, to preserve completeness, we give the option to increase the time limit each time restart is called. 

%% file: 5.experiment.tex
\section{Experiments}\label{sec:experiments}
\subsection{Setup} 
Our experiments focused on bug-finding only, and accordingly we implemented our suggested algorithms on top of \simplecar~\cite{LDPRV18,simplecar}, which is an implementation of the CAR algorithm, in its best version for bug-finding~\cite{DLPVR19}. We compared ourselves to the best public \BMC implementation (the one in \tool{ABC-BMC}~\cite{BM10}), and the best combination of CAR and BMC in~\cite{ZXLPS22}. Our evaluation was based on 438 benchmarks\footnote{Results of these benchmarks are either known to be unsafe or remain unknown.} in the Aiger~\cite{brummayer2007aiger} format from the single safety property track of the 2015~\cite{hwmcc15} and 2017~\cite{hwmcc17} Hardware Model Checking Competition (HWMCC~\cite{8102233})\footnote{These are the last two years of HWMCC using the AIGER format. Since 2019~\cite{hwmcc19}, the official format switched to a word-level format BTOR~\cite{10.1007/978-3-319-96145-3_32,brummayer2008btor}.}, which is consistent with the benchmark set of~\cite{ZXLPS22}. All the counterexamples found were successfully verified with the third-party tool \emph{aigsim} that comes with the Aiger package~\cite{aiger}. All the artifacts are available in Github~\cite{caramel}.

We ran the experiments on a cluster of Linux servers, each equipped with an Intel Xeon Gold 6132 14-core processor at 2.6 GHz and 96 GB RAM. The version of the operating system is Red Hat 4.8.5-16. For each running instance, the memory was limited to 8 GB; if not otherwise specified, the time was limited to 1 hour.

The following questions guided our evaluation:
\begin{itemize}
\item Q1: How does \emph{\local} perform when compared to the present best reordering strategy in \CAR, i.e., \emph{Intersection} + \emph{Rotation}?
\item Q2: How useful can it be to integrate \emph{\local} to the best CAR variants for bug-finding, i.e., the three presented in \cite{ZXLPS22}?
\item Q3: How does \approach perform when compared to the state-of-the-art bug-finding (unsafe checking) algorithms?
\end{itemize}

\input{5.2RQ2}
\input{5.3RQ3}

\input{5.4RQ4}

%% file: 5.2RQ2.tex
\noindent\textbf{A1. \emph{\local} VS. \emph{Intersection} + \emph{Rotation}.} 
The previous literal-ordering strategy for \CAR, namely the combination of \emph{Intersection} and \emph{Rotation} as published in~\cite{DLPVR19}, is very close to \emph{\local} when $iLimit$ is set to one (`Local-1'), except that \emph{\local} introduces a reordering inside the \UCs (see Sec.~\ref{subsection:our new reorder}). 
 Recall that in Sec.~\ref{sec:revisit} we presented empirical evidence that confirms independently of~\cite{DLPVR19} that these two strategies improve the empirical results.

\begin{figure}[!ht]
\centering
\label{Fig1}
\includegraphics[trim=0cm 0cm 0cm 0cm, clip, scale=0.13]
{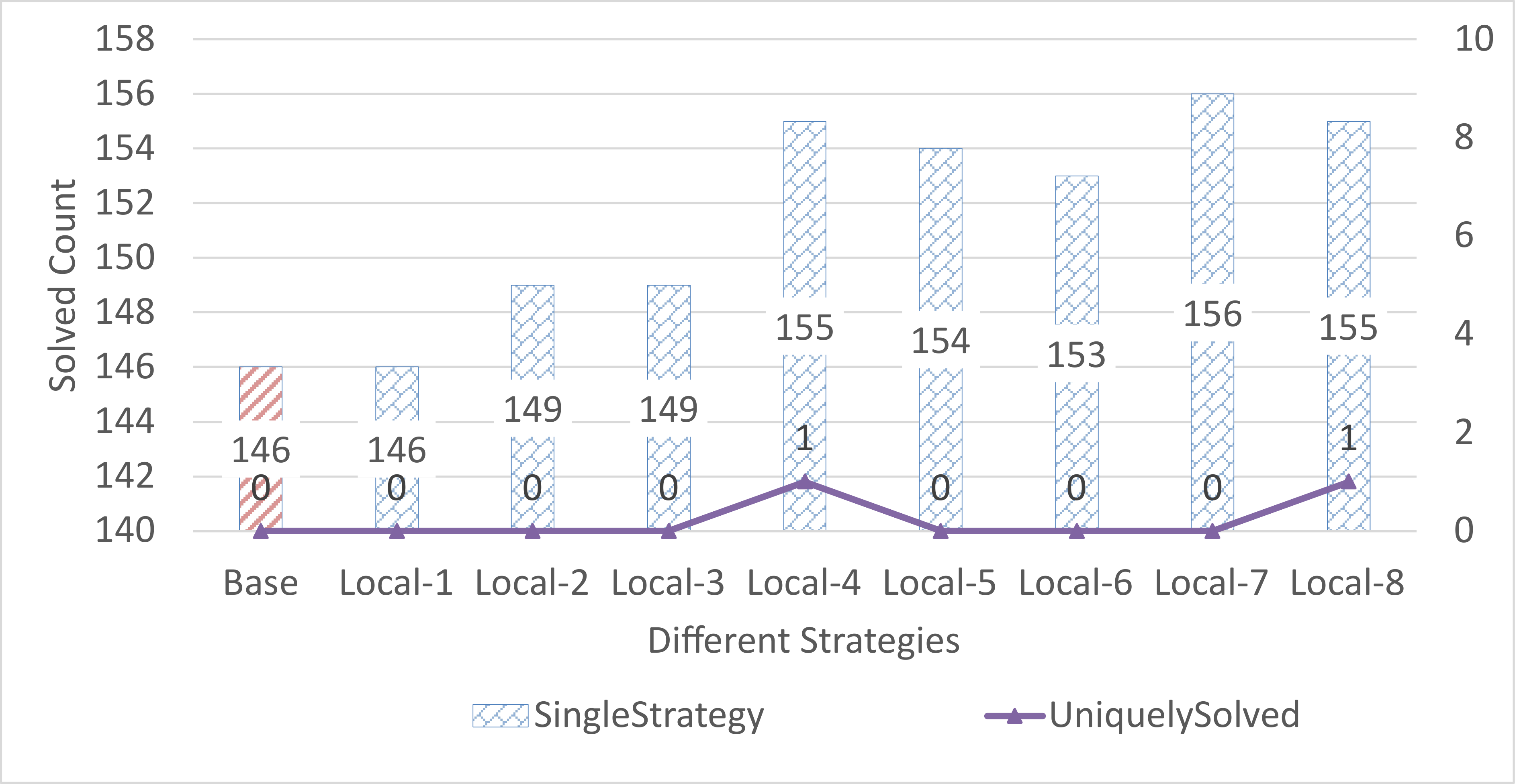}
\caption{Results on different reordering strategies, in terms of the total solved instances. In the figure, `Base' refers to the combination of \emph{Intersection} and \emph{Rotation}, `Local-$i$' ($1\leq i\leq 8$) represents the \emph{\local} strategy with $iLimit =i$.} \label{fig:local1}
\end{figure}

As is shown in Figure \ref{fig:local1}, the performance of \emph{\local} with $1 < iLimit \leq 8$ outperforms that of the base strategy. The peak performance occurs with $iLimit = 7$, which solves 156 cases in total and obtains a 7\% improvement compared to the prior best strategy (Base) in \cite{DLPVR19}.
The various strategies solve different cases, as is evident by looking at the graphs depicting the virtual best solver, with and without the base. No instance is uniquely solved by Base, indicating that a \emph{Local} portfolio can cover all the cases.  

Increasing the number of solved cases, even by a few instances, is important in light of the decades of research and development of model checkers.

\begin{figure}[!ht]
\centering
\label{Fig2}
\includegraphics[trim=0cm 0cm 0cm 0cm, clip, scale=0.13]
{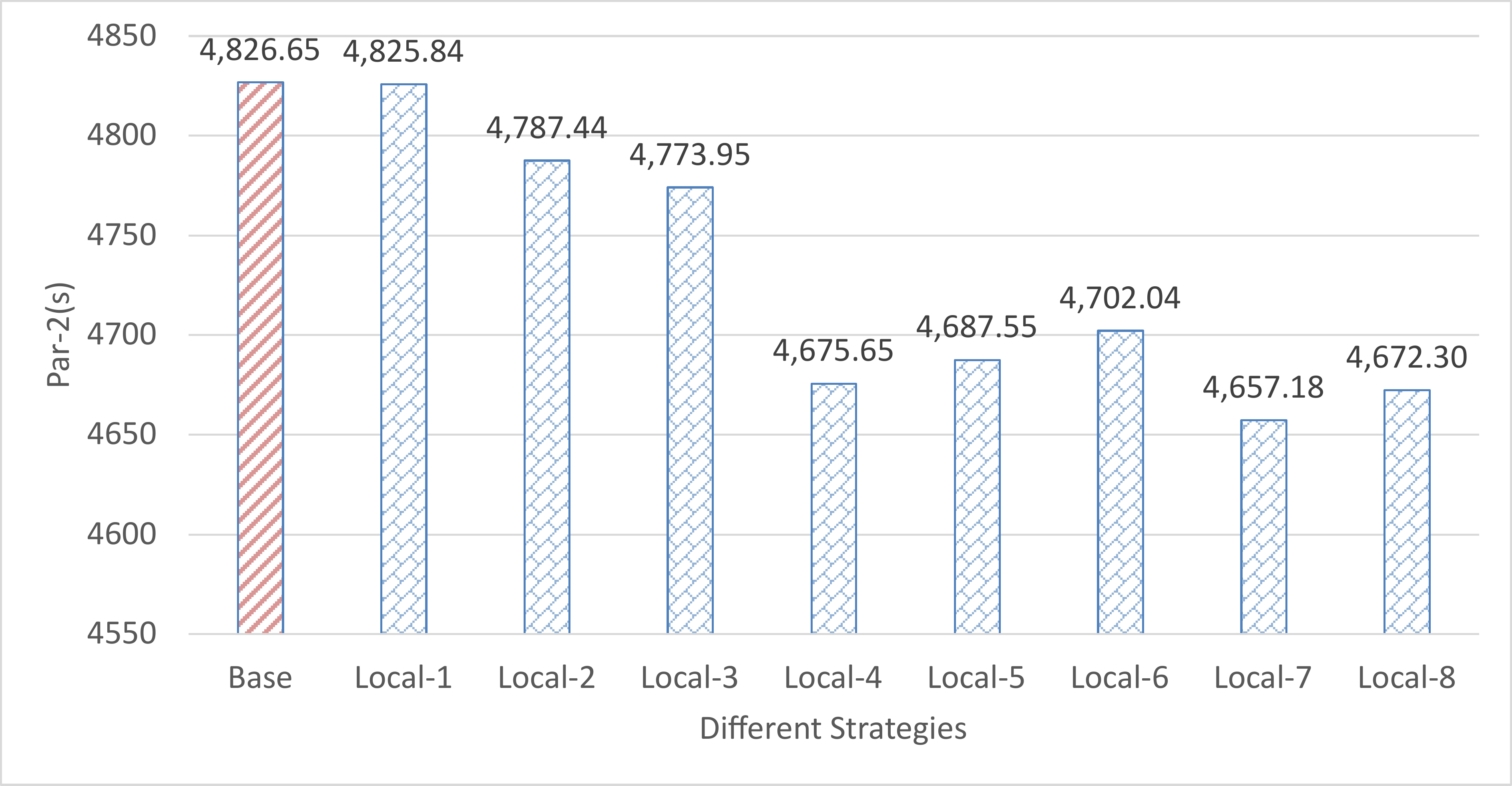}
\caption{Results on different reordering strategies, using the Par-2 score.} \label{fig:local2}
\end{figure}

A comparison of the run time of the different strategies is shown in Figure \ref{fig:local2}. We rank them using the Par-2 score, which is calculated by the average time consumption of all cases while doubling the run time of instances that timed out. This is a common factor measured in the SAT community~\cite{SATCOMP}.
The figure illustrates that \emph{\local} with $iLimit>1$ consumes less time than Base. It is also apparent that there is a correlation between the number of solved instances and the total time consumption. \emph{\local} outperforms the Base strategy on both the number and time of solved cases. A detailed pairwise comparison between the peak and Base is shown on the right. 
\begin{wrapfigure}{R}{0.5\columnwidth}	
	\includegraphics[width = 0.5\columnwidth, trim=0cm 1.0cm 0cm 2.0cm, clip]{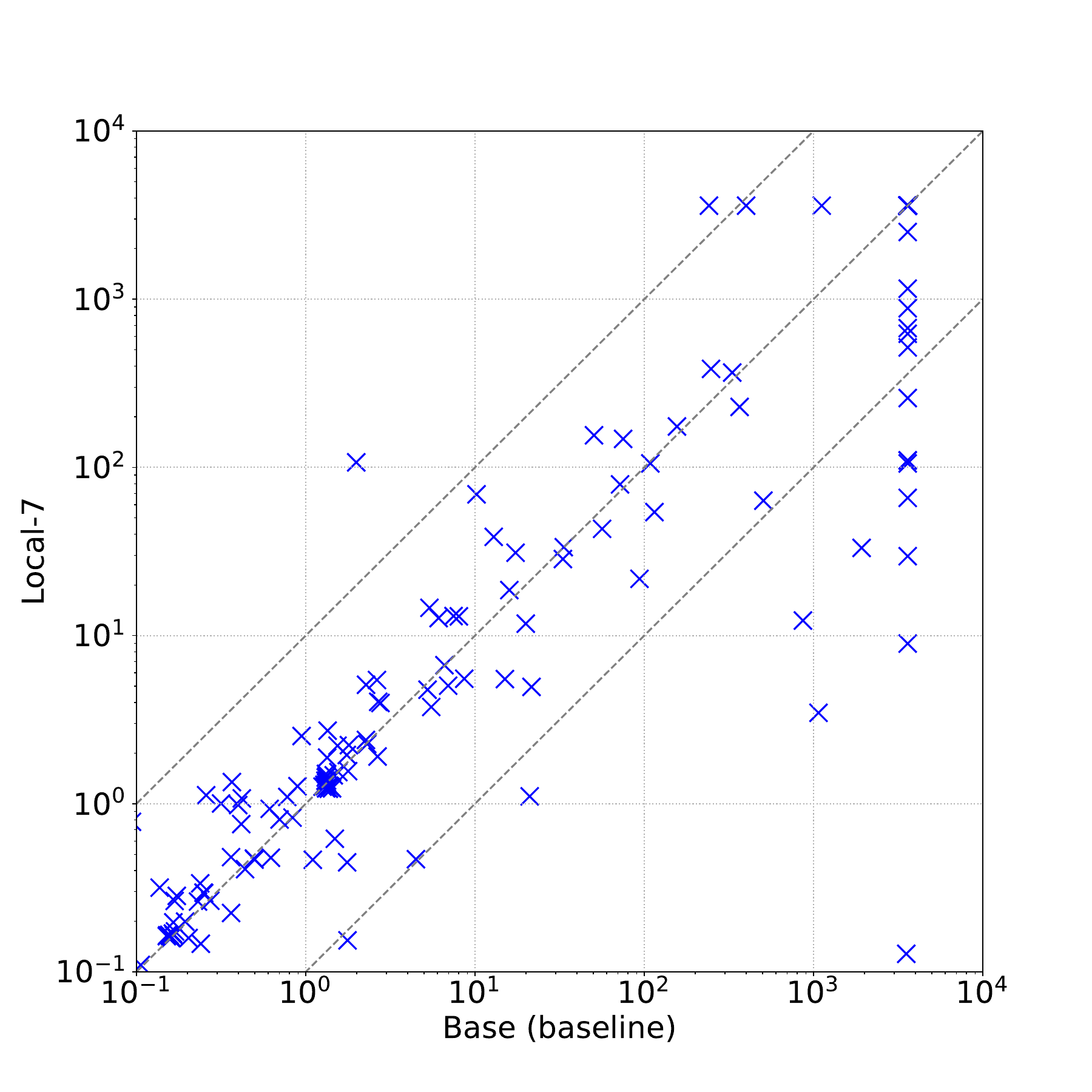}
\end{wrapfigure}

Notably, the performance of \emph{\local} with different configurations is not correlated to the value of $iLimit$. This is consistent with our discussion in Section \ref {subsection:our new reorder}, that increasing the limit does not have a monotonic effect. 

%% file: 5.3RQ3.tex

\begin{figure}
\centering
\label{Fig3}
\includegraphics[trim=0cm 0cm 0cm 0cm, clip, scale=0.13]
{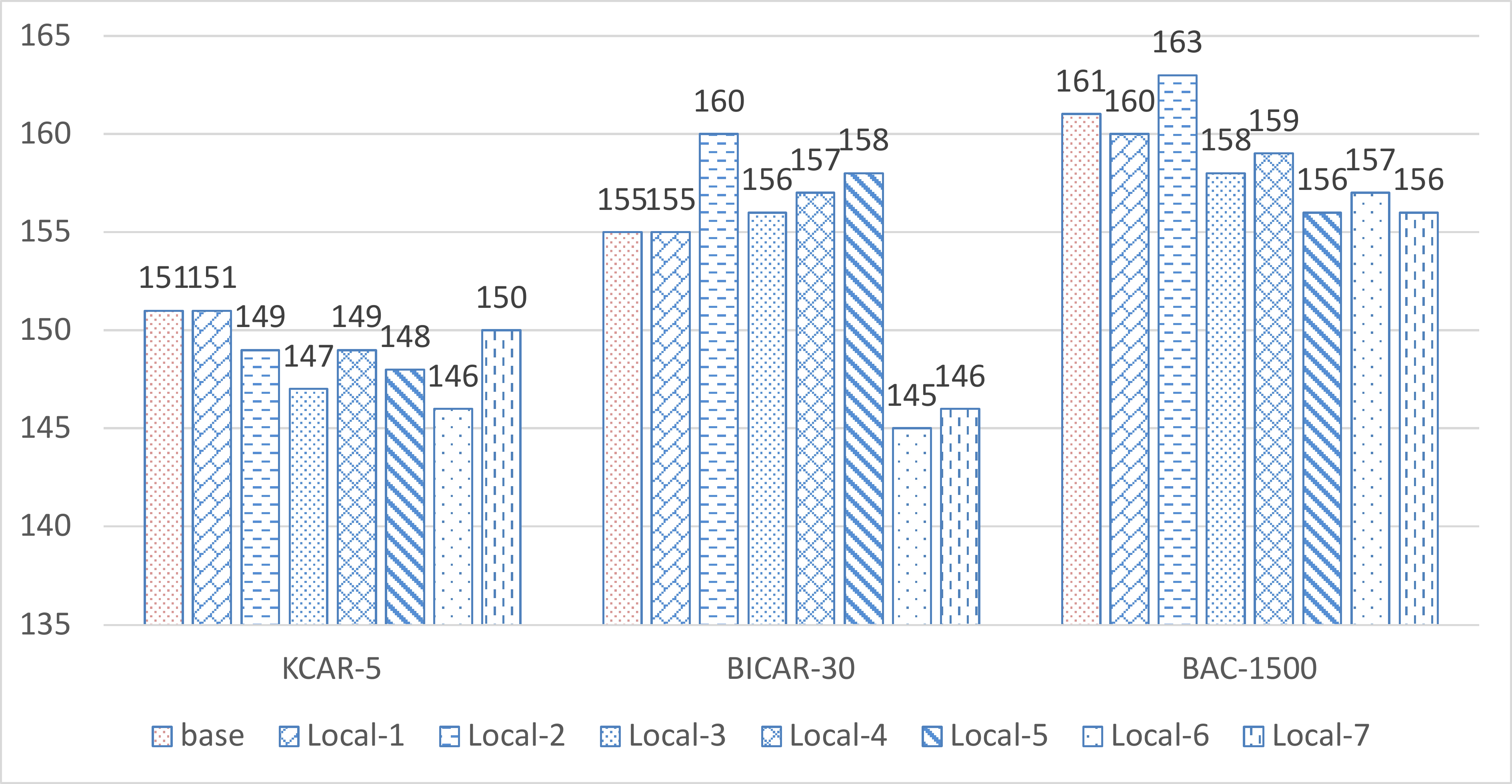}
\caption{Results of integrating \emph{\local} into KCAR-5, BICAR-30, and BAC-1500 with $iLimit$ ranging from 1 to 7.} \label{graph:fig3}
\end{figure}


\noindent\textbf{A2. The effect of \emph{\local} on the \CAR+\BMC combination of \cite{ZXLPS22}.} 
We also implemented and evaluated \emph{\local} on top of the three best combinations between \CAR and \BMC, i.e., BAC-1500, BICAR-30, and KCAR-5, from \cite{ZXLPS22}. The results are shown in Figure \ref{graph:fig3}.
Generally speaking, \emph{\local} can be helpful for BICAR-30 and BAC-1500 to solve more instances when $iLimit=2$ (2 and 5 more instances, respectively), though it seems to be detrimental in other cases.

%% file: 5.4RQ4.tex


\begin{figure}[!ht]
\centering
\setlength{\abovecaptionskip}{0cm} 
\includegraphics[trim=0cm 0cm 0cm 0cm, clip, scale=0.13]
{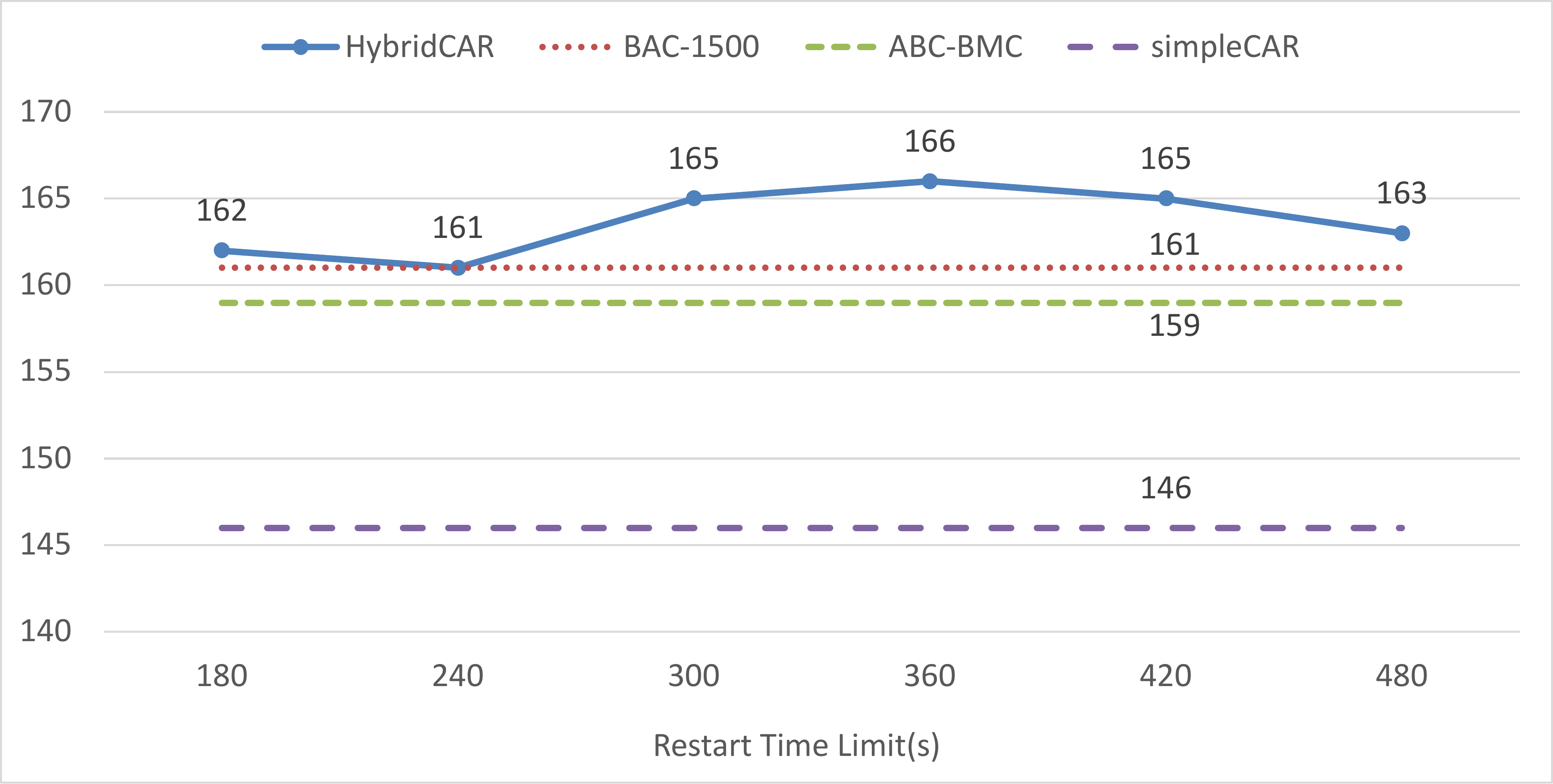}
\caption{Comparison on \approach with different restarting limits to \tool{BAC-1500}, \tool{ABC-BMC}, as well as the original \simplecar. Time limit is 1 hour. The X-axis represents different restarting limits.} \label{graph:different_hybrid_1h}
\end{figure}

\begin{figure}[!ht]
\centering
\setlength{\abovecaptionskip}{0cm} 
\includegraphics[trim=0cm 0cm 0cm 0cm, clip, scale=0.13]
{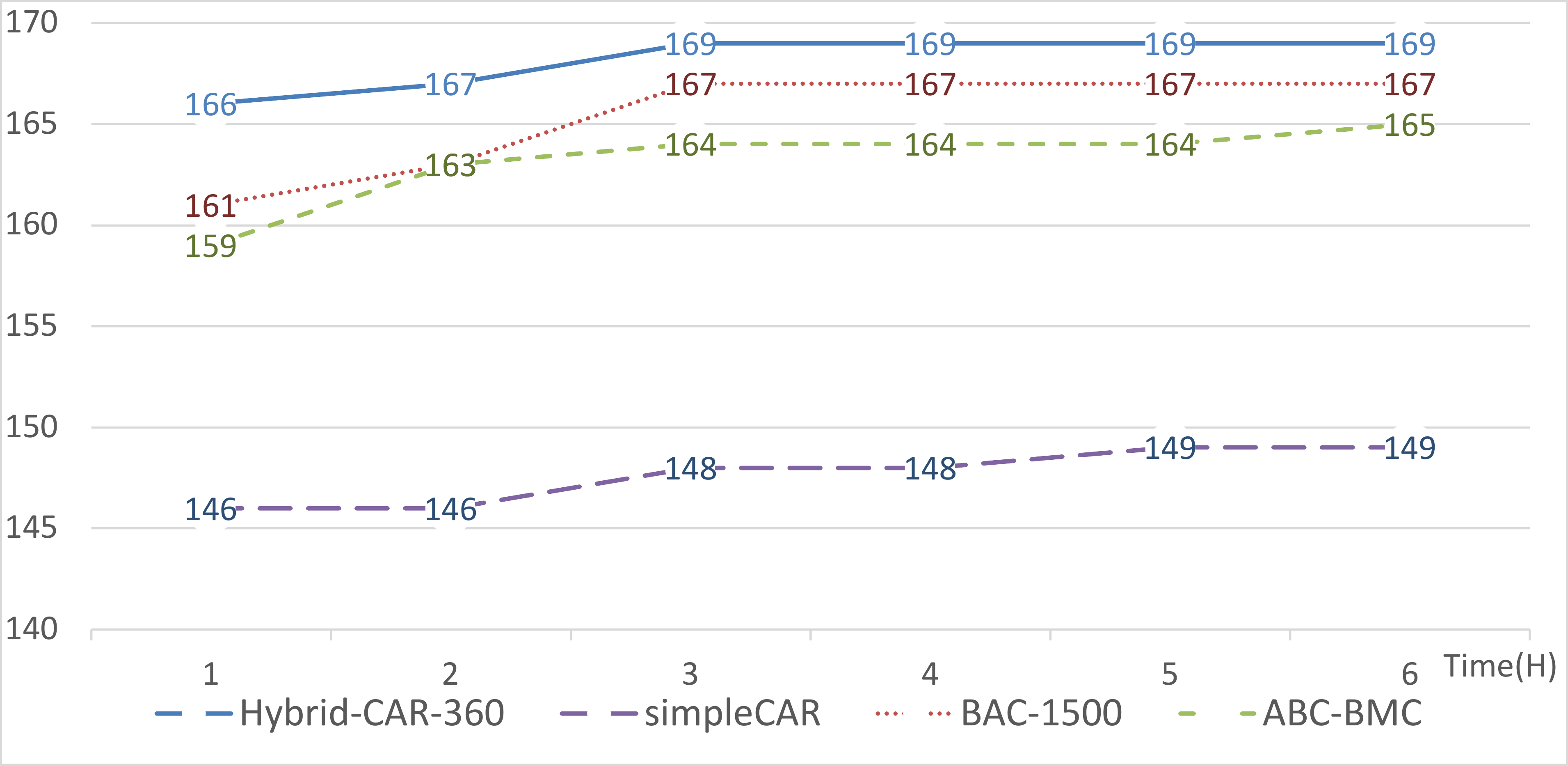}
\caption{Comparison on \approach with different restarting limits to the original \simplecar, \tool{ABC-BMC}, and \tool{BAC-1500}, the latter two of which are shown to be the state of the art in \cite{ZXLPS22}. In the figure, \approach-360 refers to \approach with the restart limit set to 360 seconds. Timeout is up to 6 hours. The X-axis represents CPU running time.} \label{graph:different_hybrid_6h}
\end{figure}

\noindent\textbf{A3. \approach VS. state-of-the-art bug-finding algorithms.} 
We compared \approach to the original \simplecar, \tool{BAC-1500} (the best solution shown in \cite{ZXLPS22}) and \tool{ABC-BMC} on bug finding\footnote{While \PDR is good at proving safe, it is not as good in finding bugs. The best implementations of \PDR, to our knowledge, namely \tool{abc-pdr} and \tool{nuxmv-ic3}, cannot solve more than 140 cases within the time limit. So do other variants such as \tool{Avy} \cite{VG14,VVGG19} and \tool{QUIP} \cite{GI15}. For this reason they are not included in the comparison.}. To fully evaluate the scalability of these different approaches, we ran the experiments with two separate time limits: 1 hour and 6 hours. 
The corresponding results are shown in Figure \ref{graph:different_hybrid_1h} (1 hour) and Figure \ref{graph:different_hybrid_6h} (6 hours). 

It turns out that regardless of the time limit for restart, \approach performs better than the competitors. In the one-hour setting, the best version of \approach, in which the restart is invoked every 6 minutes, solves 166 cases in total, which is seven more than that solved by \tool{ABC-BMC}, and 20 more than that solved by the original \simplecar. 
In the 6-hour setting, this version of \approach solves 169 cases, which is 20 more than the original \simplecar, and outperforms the competition.  Note that \approach solves more instances in just one hour (166) than that solved by \tool{ABC-BMC} in 6 hours (165).


\begin{figure}
    \centering
    \setlength{\abovecaptionskip}{0cm}
    \includegraphics[trim=0cm 0cm 0cm 0cm, clip, scale=0.11]{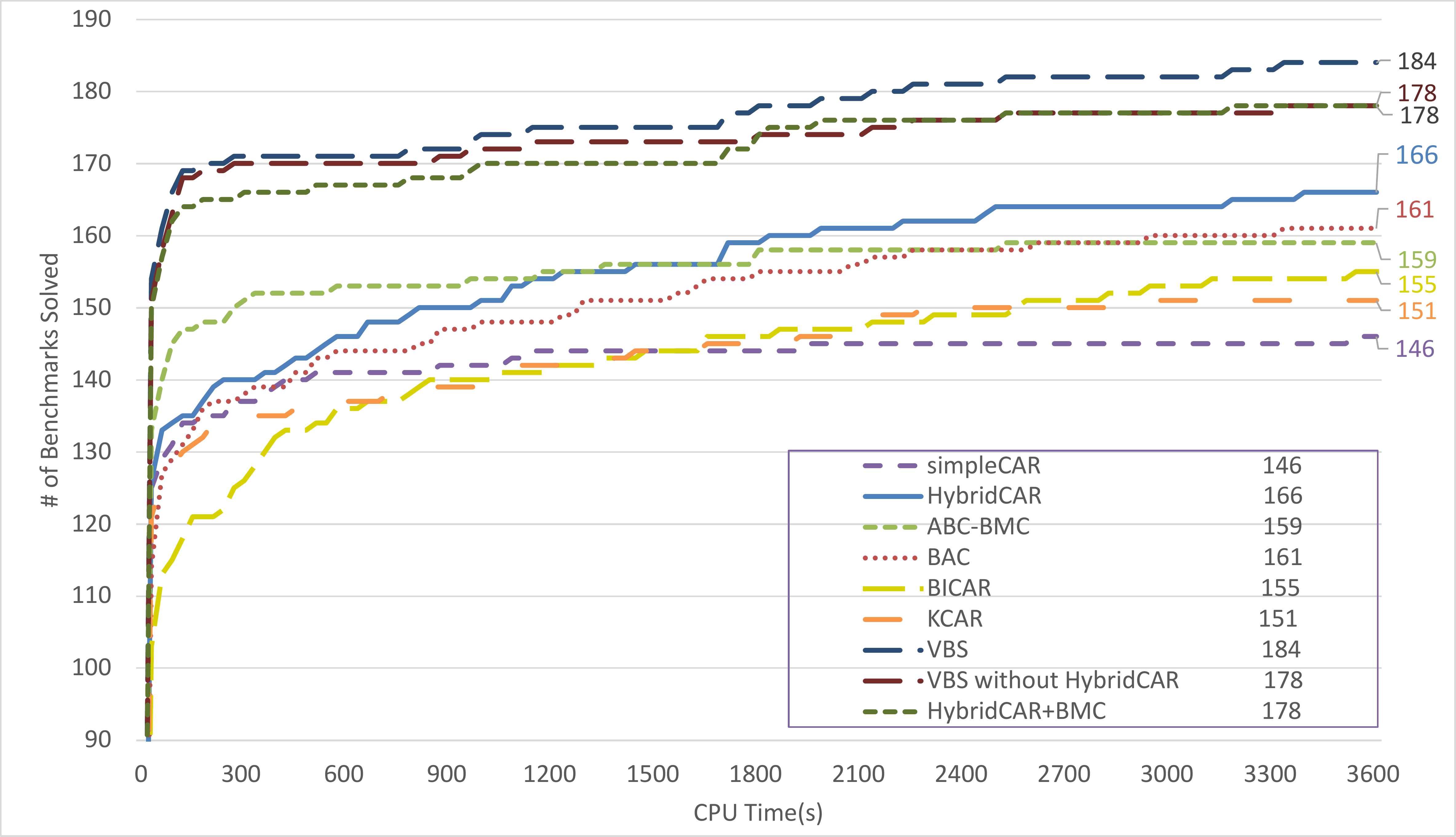}
    \caption{Comparison of run-time performance among different model checkers. VBS represents the virtual best, i.e., parallel running all and taking the best.} \label{graph:10}
\end{figure}

Figure~\ref{graph:10} also includes the comparison among the best version of \approach and the other two combinations of \BMC and \CAR presented in \cite{ZXLPS22}, i.e., BICAR and KCAR. The timeout here is one hour. 
\approach performs better than all the other methods. In terms of the number of solved unsafe instances, \approach is 166, followed by BAC (161), ABC-BMC (159), BICAR (155), KCAR (151), and the original \CAR (146). 
In particular, \approach can solve six \emph{unique} benchmarks, i.e., benchmarks that cannot be solved by all the other methods. 

\begin{table}
\renewcommand{\arraystretch}{1.4}
\caption{Uniquely Solved Instances for each algorithm.}
\label{tab:hybrid-3}
\scriptsize
\centering
\begin{tabular*}{\hsize}{l c c c}
    \hline
    &\thead{UNIQUELY\\SOLVED} & \thead{UNIQUELY \\SOLVED\\(COMPARED\\ TO ABC-BMC) }  & \thead{UNIQUELY\\ SOLVED\\(COMPARED \\TO BAC)}\\
    \hline
    simpleCAR & 0& 13 & 4\\
    \textbf{\approach} &\textbf{6} & \textbf{19} & 11\\
    BAC & 4 & 17 & 0\\
    ABC-BMC & 5 & 0 & \textbf{15}\\
    BICAR & 0 & 5 &  10\\
    KCAR & 1 & 10 & 6\\
    \hline
\end{tabular*}
\end{table}

\begin{figure}
    \centering
    \setlength{\abovecaptionskip}{0cm}
    \includegraphics[trim=0cm 0cm 0cm 0cm, clip, width=\columnwidth]{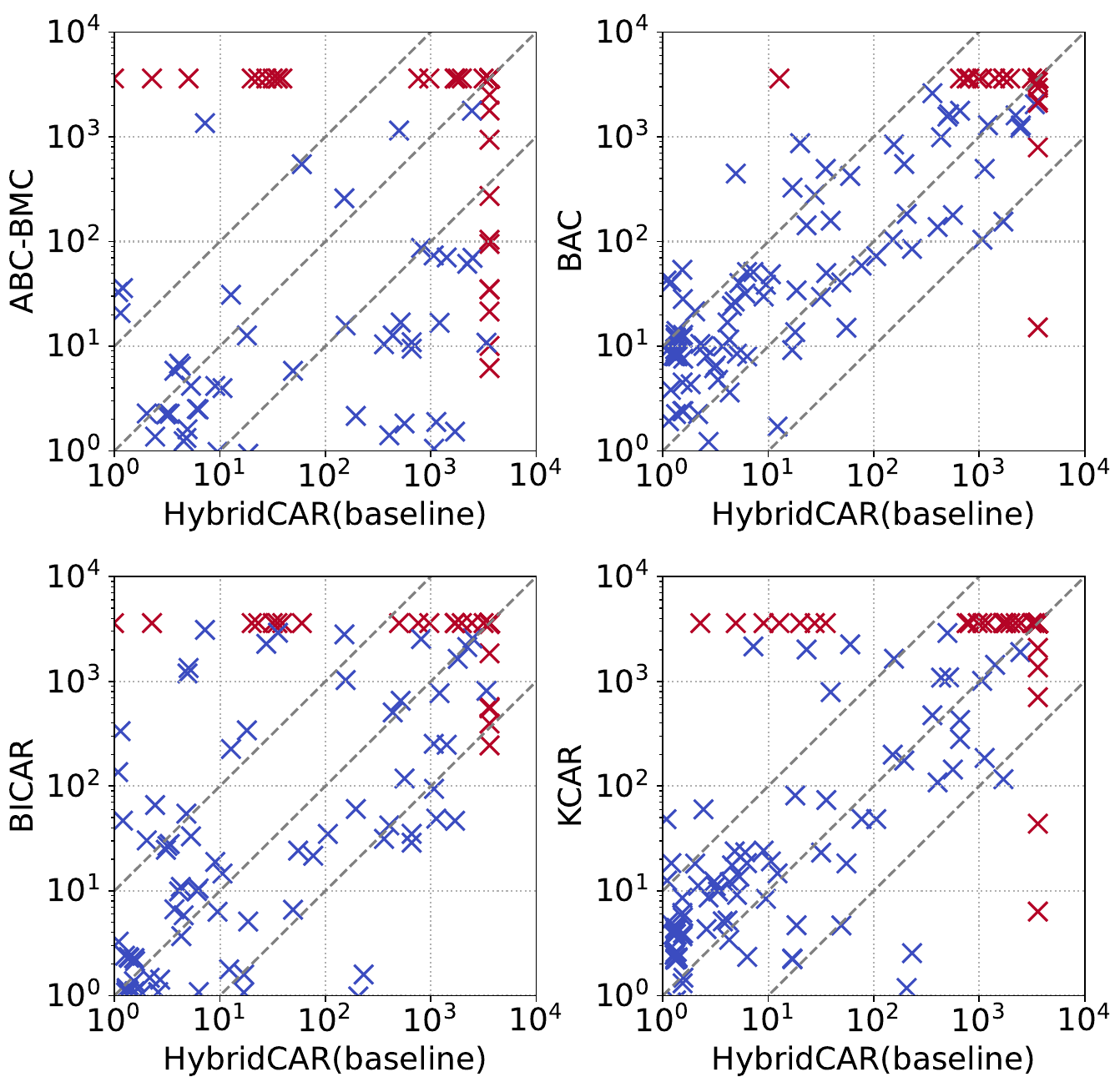}
    \caption{Pairwise Comparison of \approach and competitors. Timeout instances in either are marked in red. } \label{graph:pairwise}
\end{figure}

Table \ref{tab:hybrid-3} shows the uniquely solved instances of each technique  (i.e. that no other tool can solve), and, in parenthesis, in comparison to \tool{ABC-BMC} and \tool{BAC}, e.g., \approach solves 19 and 11 cases that cannot be solved by these two tools, respectively. Moreover, we note that a portfolio of only \approach and \tool{ABC-BMC} can solve 178 instances, almost reaching the virtual best results (184) that a portfolio of all these algorithms can solve. A detailed pairwise comparison is shown in Fig.~\ref{graph:pairwise}.

%% file: 6.conclusion.tex
\section{Conclusion}\label{sec:conclusion}
In this paper, we revisited the assumption literal ordering strategies presented in~\cite{DLPVR19}. We hypothesized that \emph{Intersection} works because of what we call \emph{core locality}, which means that similar cores help the SAT solver find proofs faster. Our empirical data, as we showed, supports this claim.  Both \emph{Intersection} and \emph{Rotation} determine only a part of the literal order, hence the order of most of the assumptions is left arbitrary. Our improved strategy, \local (Sec.~\ref{subsection:our new reorder}), generalizes \emph{Intersection} and orders a larger part of the assumptions sequence, while improving the core locality. Together with prioritizing \emph{conflict literals} (Sec.~\ref{sec:conflictliterals}) they shorten rather significantly the time it takes the SAT solver to find proofs. 
We also presented a hybrid approach called \approach (Sec.~\ref{sec:approach}), which switches between different configurations of \local during run time, while resetting the $U$ sequences. 
Our results show that these strategies perform better on average than the reordering strategies of~\cite{DLPVR19} and also better than the various integrations of \CAR with \BMC~\cite{ZXLPS22}. In particular, \approach is able to outperform all bug-finding model-checking algorithms off-the-shelf. It is left for future work to try these strategies on \PDR.

%% file: bio.tex
\begin{IEEEbiography}
[{\includegraphics[width=1in,height=1.25in,clip,keepaspectratio]{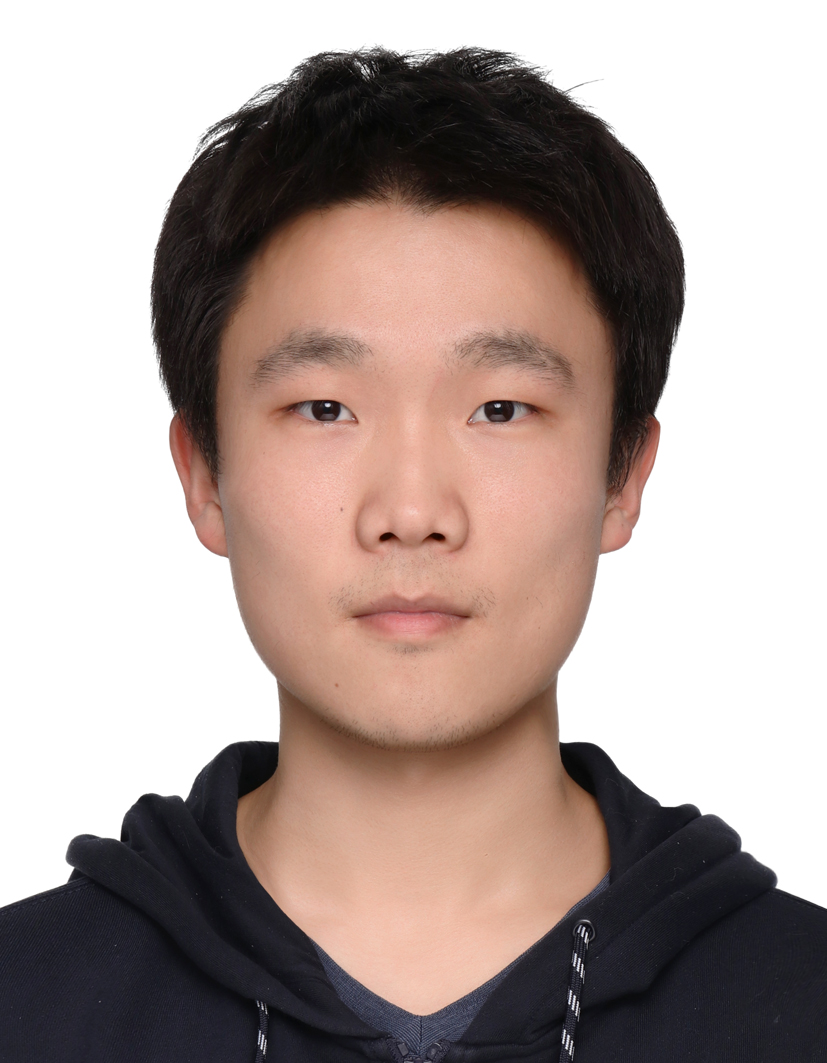}}]{Yibo Dong} 
received the B.S. degree from Shanghai Jiao Tong University, Shanghai, China, in 2021. He is pursuing an M.S. degree with the Software Engineering Institute at East China Normal University, Shanghai.
His main research interest lies in formal verification, especially model checking.
\end{IEEEbiography}

\begin{IEEEbiography}
[{\includegraphics[width=1in,height=1.25in,clip,keepaspectratio]{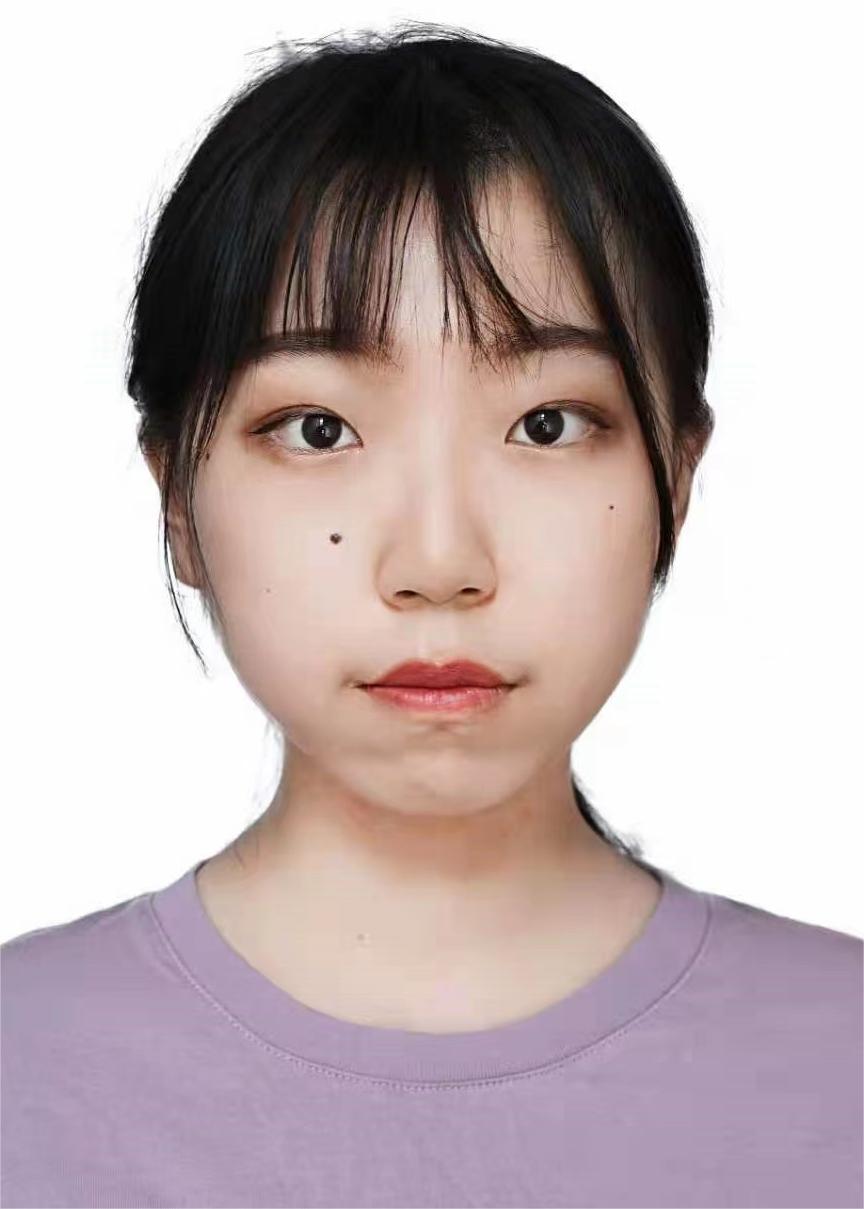}}]{Yu Chen}
received her B.S. degree from Shanghai University, Shanghai, China, in 2020 and her M.S. degree from East China Normal University, Shanghai, China, in 2023. She is currently a teaching assistant at Chuzhou University, Anhui, China.
Her main research interest lies in temporal logic and model checking.
\end{IEEEbiography}

\begin{IEEEbiography}
[{\includegraphics[width=1in,height=1.25in,clip,keepaspectratio]{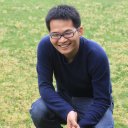}}]{Jianwen Li}
He received his Ph.D. degree from the Software Engineering Institute, East China Normal University, Shanghai, China, in 2014. He is currently a Research Professor at the Software Engineering Institute, East China Normal University. His research interests include formal verification, logic and automata theory.
\end{IEEEbiography}

\begin{IEEEbiography}
[{\includegraphics[width=1in,height=1.25in,clip,keepaspectratio]{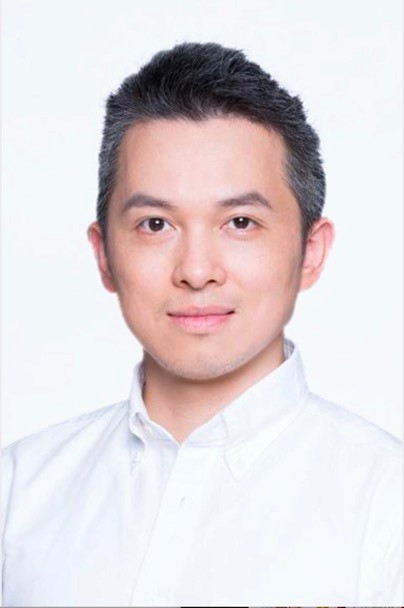}}]{Geguang Pu}
received his B.S. degree in mathematics from Wuhan University, Wuhan, China, in 2000, and his Ph.D. degree in mathematics from Peking University, Beijing, China, in 2005. He is currently a Professor at the Software Engineering Institute, East China Normal University, Shanghai, China. He has published over 100 publications on software engineering and system verification, including ICSE, FSE, ASE, and CAV. His research interests include program testing and reliable AI systems. Prof. Pu served as a PC member for more than 20 international conference committees.
\end{IEEEbiography}

\begin{IEEEbiography}
[{\includegraphics[width=1in,height=1.25in,clip,keepaspectratio]{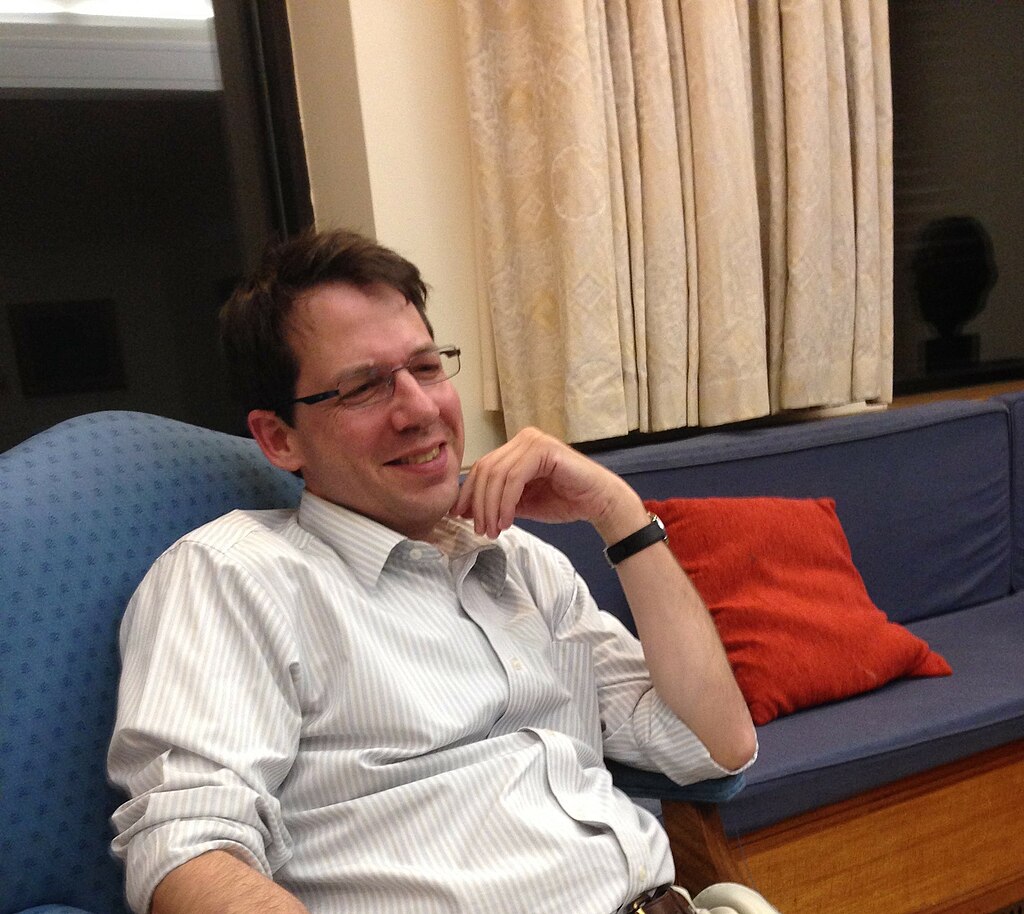}}]{Ofer Strichman} 
Prof. Ofer Strichman earned his PhD in 2001 from the Weizmann Institute, where he worked, under the supervision of Amir Pnueli, on translation validation for compilers, Bounded Model Checking, and other topics in formal verification. He then was a post-doc in Carnegie Mellon University in Ed Clark’s group, where he mostly worked on model-checking, learning, predicate abstraction and decision procedures.
Prof. Strichman published two books: “Decision procedures – an algorithmic point of view” together with Daniel Kroening, and “Efficient decision procedures for validation”, edited two others and coauthored more than 100 peer-reviewed articles, mostly in formal verification and SAT. In the SAT field he is mostly known for his contributions to linear-time proof manipulations, exploiting symmetries and incremental satisfiability. In formal verification he is mostly known for his invention of ‘regression verification’ and various decision procedures, mostly for equalities with uninterpreted functions.

Prof. Strichman won the 2021 CAV award “for pioneering contributions to the foundations of the theory and practice of satisfiability modulo theories (SMT)”.
\end{IEEEbiography}